\documentstyle[psfig,10pt,aaspp4]{article}
\def\kms{km~s$^{-1}$}

\def\be{\begin{equation}}
\def\ee{\end{equation}}
\def\about{$\sim$}
\def\etal{{\it et al.}}
\def\h{$h^{-1}$}
\def\HI{\ion{H}{1}}

\def\deg{$^{\circ}$}

\def\r83{$r_{83}$}

\begin{document}
\hskip 3.5in{\hskip 10pt \date{1 August 1998}}
\title{SEEKING THE LOCAL CONVERGENCE DEPTH. V. TULLY-FISHER PECULIAR VELOCITIES FOR 52 ABELL CLUSTERS.}
 
\author {DANIEL A. DALE,\altaffilmark{1} RICCARDO GIOVANELLI, AND MARTHA P. HAYNES,}\affil{Center for Radiophysics and Space Research and National Astronomy and Ionosphere Center, Cornell University, Ithaca, NY 14853}

\author {LUIS E. CAMPUSANO}
\affil{Observatorio Astron\'{o}mico Cerro Cal\'{a}n, Departamento de 
Astronom\'{\i}a, Universidad de Chile, Casilla 36-D, Santiago, Chile}

\author {EDUARDO HARDY}
\affil{National Radio Astronomy Observatory, Casilla 36-D, Santiago, Chile}

\altaffiltext{1}{Now at IPAC, California Institute of Technology 100-22, Pasadena, CA 91125}

\begin{abstract}
We have obtained $I$ band Tully-Fisher (TF) measurements for 522 late-type galaxies in the fields of 52 rich Abell clusters distributed throughout the sky between \about 50 and 200\h\ Mpc.  Here we estimate corrections to the data for various forms of observational bias, most notably Malmquist and cluster population incompleteness bias.  The bias-corrected data are applied to the construction of an $I$ band TF template, resulting in a relation with a dispersion of 0.38 magnitudes and a kinematical zero-point accurate to 0.02 magnitudes.  This represents the most accurate TF template relation currently available.  Individual cluster TF relations are referred to the average template relation to compute cluster peculiar motions.  The line-of-sight dispersion in the peculiar motions is $341\pm93$ \kms, in general agreement with that found for the cluster sample of Giovanelli and coworkers.
\end{abstract}

\keywords{galaxies: distances and redshifts --- cosmology: 
observations; distance scale}

%\section {INTRODUCTION}
\section {Introduction}
\label{sec:intro}
Deviations from smooth Hubble flow arise as a result of large scale density fluctuations.  Qualitatively, a ``convergence depth'' is the distance out to which significant contributions to a galaxy's peculiar motion are made.  In linear theory, the fraction of the peculiar velocity of a galaxy, contributed by the mass distribution within a radius $R$, can be written as
\be
{\bf V}(R) \simeq {H_{\circ} \Omega^{0.6}_{\circ} \over {4 \pi}} \int \delta( {\bf r}) {\hat {\bf r} \over r^2} W(R) d^3{\bf r},
\label{eq:Vpec}
\ee
where the Hubble constant $H_\circ= 100h$ \kms\ Mpc$^{-1}$,  $\Omega_\circ$ is the cosmological matter density parameter, $\delta$ is the mass overdensity, and $W(R)$ is a window function of width $R$ (a Gaussian or top-hat, for example) centered at $r$=0 (Peebles 1993).  If the distribution of matter approximates homogeneity on larger scales, then contributions to the peculiar velocity will eventually taper off with increasing distance.

Observational estimations of the local convergence depth are facilitated through measurements of the reflex motion of the Local Group with respect to spherical shells of increasing radii.  The first such program was carried out by Rubin \etal\ (1976) on an all-sky sample of 96 Sc galaxies between redshifts of 3500 and 6500 \kms.  They reported the motion of the Local Group with respect to the shell of galaxies was 454$\pm$125 \kms\ towards $(l,b)=(163^\circ,-11^\circ)$, significantly different from the apex of the CMB dipole, suggesting a large bulk flow for the shell and a convergence depth significantly larger than \about 50\h\ Mpc.  Subsequent work in this area has yielded conflicting results.  

The situation can be grossly expressed into two main views.  A picture of a {\it small} convergence depth (c$z < 5000$ \kms) was first suggested by Tammann \& Sandage (1985) and successively by Dressler \etal\ (1987) and Lynden--Bell \etal\ (1988).  This picture is also supported by early measurements of the distribution of IRAS and optical galaxies (Lahav, Lynden-Bell \& Rowan-Robinson 1988; Lynden-Bell, Lahav \& Burstein 1989; Strauss \etal\ 1992; Hudson 1993).  It should be mentioned, however, that some IRAS dipoles suggest significant contributions to the Local Group motion arise from the distribution of objects between 5000 and 10,000 \kms\ (e.g. Rowan-Robinson \etal\ 1990).  The latter studies profit from flux dipoles or a peculiar velocity field that is solved for iteratively, initially supposing that recessional velocities are indicative of distance and that light traces mass.  The above observational programs that utilize this technique find 80\% or more of the Local Group motion derives from matter within \about 50\h\ Mpc. 

Proposals for a {\it large} convergence depth (c$z > 10,000$ \kms; Scaramella, Vettolani \& Zamorani 1994; Tini-Brunozzi \etal\ 1995; Branchini \& Plionis 1996; Plionis \& Kolokotronis 1998), mainly resulting from a similar analysis of the distribution of clusters of galaxies, suggest that much of our motion may be produced by mass concentrations as far as \about 13,000 \kms\ (e.g. the ``Shapley Supercluster'').  The latter view is corroborated by reports of bulk motion in the local Universe (within 6000 \kms) via Tully--Fisher (TF) measurements by Willick (1990), Courteau \etal\ (1993) and Mathewson, Ford \& Buckhorn (1992), and by the recent analysis of the dipole of cluster brightest ellipticals by Lauer and Postman (1994, henceforth LP).  The LP claim, based on the peculiar motions of all Abell clusters to 15,000 \kms, suggests the local volume of space within \about 100\h\ Mpc is traveling towards $(l,b)=(343^\circ,+52^\circ)$ at 689 \kms.  The disagreement between these competing views of the convergence depth scale is wide, with important implications for cosmological models, which are generally unable to accommodate bulk flows with scales as implied by the large convergence depth camp (e.g. Gramann \etal\ 1995).  

Fresh work from Hudson \etal\ (1999) and Willick (1999) challenge the direction of the LP bulk flow vector.  Hudson \etal\ employ the fundamental plane relation for some 700 ellipticals in 56 clusters (3000 $\lesssim$ c$z \lesssim$ 14,000 \kms) to find a bulk flow of 630$\pm$200 \kms\ in the direction $(l,b)=(260^\circ,-1^\circ)$.  A different bulk motion of \about 700$\pm$250 \kms, towards $(l,b)=(272^\circ,20^\circ)$ originating from TF measurements in 172 cluster galaxies (9000 $\lesssim$ c$z \lesssim$ 13,000 \kms) and 72 other galaxies (c$z$ $<$ 30,000 \kms), is claimed by Willick.  Such large bulk flows are not consistent with other recent observational work.  Riess, Press \& Kirshner (1995) used 13 SN Ia observations and found evidence for a small local bulk flow.  In a contribution based on TF distances of about 2000 galaxies within 9500 \kms, Giovanelli \etal\ (1998a) similarly report evidence against the existence of a large scale local flow, and support for a relatively small convergence depth.  Many of the aforementioned studies cannot convincingly exclude the existence of large-scale bulk flows, either because their sampling is too sparse (Watkins \& Feldman 1995) or of limited depth.  This is particularly important in view of the claims that asymptotic convergence of the Local Group reflex motion may only be reached at distances well in excess of 10,000 \kms\ (Scaramella, Vettolani \& Zamorani 1994, Tini Brunozzi \etal\ 1995, and Branchini \& Plionis 1996).  This work aims at the direct determination of the TF relation for an all-sky cluster set extending to distances exceeding the highest reported values of the convergence depth.

In addition to ensuring that our TF relation is valid for usefully large distances, a second issue of concern regards the amplitude of systematic errors in the TF template.  The Giovanelli \etal\ (1998a) sample of peculiar velocities obtained using the TF method is referred to a template relation based on the SCI sample, a collection of 782 TF measurements in the fields of 24 separate clusters between 1000 and 9000 \kms\ (Giovanelli \etal\ 1997a,b; hereafter G97a,b).  Though the relative proximity of the SCI allows a broad stretch of observable galactic properties and thus makes it an ideal sample to study several characteristics of the TF template relation, it has limits as to how accurately the relation's zero point can be pinned down.  A larger and deeper cluster sample has two main advantages.  First, the increased number of clusters reduces the impact of statistical ``shot noise.''  In addition, since the magnitude offset produced in the TF diagram by a given peculiar velocity decreases with the target distance, the scatter produced by peculiar velocities of distant clusters about the template relation is reduced; thus they are better suited to determining the template relation's zero point.  Here we present the results of a program designed to probe the large-scale peculiar velocity field to \about 200\h\ Mpc, consisting of spectroscopic and photometric data for an all-sky sample of 522 galaxies from the fields of 52 clusters between \about 50 and 200\h\ Mpc (hereafter the `SCII' sample).\footnote{The SCI and SCII are complementary samples of cluster TF data.  The SFI is a completely independent sample of TF data for some 2000 field galaxies.  Details on the SCI and SFI samples can be found in Giovanelli \etal\ (1998a,b) and references therein.}

The overall scatter of the $I$ band TF relation of about one-third of a magnitude translates to an uncertainty of 15\% in redshift-independent distance measurements.  This means that for an individual galaxy at, say 15,000 \kms, the method is able to predict the distance to within 2250 \kms.  This value is considerably larger than the typical value of peculiar velocities, which are of order 500 \kms\ or less.  Even if 50 objects per cluster were to be measured, the distance of a cluster at 15,000 \kms\ would not be characterized to better than $2250/\sqrt{50} \sim 300$ \kms; our measurements include many fewer galaxies per cluster field, typically 10, some of which turned out to be cluster members.  The main purpose of this study was thus {\it not} the determination of accurate {\it individual} peculiar velocities of remote clusters, but rather to combine measurements in many clusters to obtain a global solution for the dipole of the velocity field.

This is the fifth paper in our series on the local convergence depth.  The observational data are presented in Dale \etal\ (1997, 1998, 1999; Papers I, II, and IV; the data can be obtained by contacting the first author).  The core result of this work, that the local dipole flow to about 200\h\ Mpc is consistent with a null bulk motion, is described in Dale \etal\ (1999; Paper III).  Details of the sample and its selection are covered in Section 2 of this work, while Section 3 covers the construction of the universal TF template relation.  Results for the peculiar velocity sample are given in Section 4, and we summarize our findings in Section 5.

%\section{SAMPLE SELECTION}
\section{Sample Selection}

Clusters of galaxies are used as increasingly pliant tools in cosmology.  For our particular concern, the cluster peculiar velocity distribution reliably matches that of the underlying smoothed matter's velocity distribution (Bahcall \etal\ 1994a,b; Gramann \etal\ 1995).  Furthermore, though observations of clusters of galaxies only sparsely sample the large scale velocity field, they can do so more accurately than individual galaxies can.  This advantage arises because measurements of many galaxies {\it within a single bound system} can be made.  As peculiar velocities are driven by gravitationally growing density fluctuations, precise comparisons of large scale peculiar velocities can be made with those predicted by cosmological theories (Watkins \& Feldman 1995; Cen \etal\ 1994; Feldman \& Watkins 1994; Strauss \etal\ 1995; Croft \& Efstathiou 1994; Borgani \etal\ 1997). 

\subsection{Sample Definition}
\label{sec:def}

Clusters of galaxies are practical TF targets.  Their outskirts are well populated by spiral galaxies, thus a small number of wide-field images with a modest-sized telescope can effectively ``map'' a cluster, and they will likely contain numerous TF candidates.

A second consideration that suggests the use of clusters in TF experiments is the determination of the TF template.  As will be outlined in Section \ref{sec:calibration}, the template is preferentially measured within a cluster, or determined as an average of templates from measurements in separate clusters.  The template slope is best measured if a broad dynamic range of TF parameters can be observed, preferentially obtained from a relatively nearby sample.  In contrast, the kinematical zero point is better characterized at higher redshifts, where cosmic peculiar motions play a smaller role in shifting objects away from the template.  Our motive of improving the TF zero point accuracy plays an important role in characterizing the redshift distribution of our sample.   

A third advantage to using clusters of galaxies in TF work is the \about $\sqrt{N}$ increase in statistical accuracy per cluster they afford if $N$ TF measurements per cluster are available -- the estimate of a system's peculiar motion is more accurate when information from multiple objects is used, the distance differential between galaxies in a cluster being negligible for our sample.  However, if an all-sky survey of peculiar motions is the goal, concentrating observations to a single cluster or to a small number of clusters parallels an increase in overall sample sparseness.  The choice between a densely sampled volume of low accuracy peculiar velocities, and a sparse sample of accurate peculiar velocities ultimately depends on the particular goals of the study.  

%\subsection{Monte Carlo Simulations of Sample Selection}
We early on investigated what sample characteristics would facilitate the most accurate determination of the TF zero point and the local bulk flow.  To ascertain the optimal distribution of clusters and number of galaxies per cluster to observe, we relied on numerical simulations.  The only limitation enforced was the total number of objects that could be observed, for target accuracy and a reasonable project timescale.  A wide range in the number of clusters and the number of objects per cluster was explored.  A variety of models of the peculiar velocity field (multi-attractor, linear bulk flows, quiescent, etc.) was imposed, and different models for the shape of the Zone of Avoidance (ZoA) were considered.  These studies profited from numerical simulations of various types of cold dark matter models kindly provide by S. Borgani (see Paper III for further details).  TF errors were assigned using random Gaussian deviates of the scatter distribution in G97b.  The results, which took into consideration that clusters of galaxies have positions that are correlated in space (and therefore a random choice of $N$ clusters does not guarantee that $N$ independent points are sampled), suggested that about 50 clusters of galaxies needed to be observed, with measurements of at least 7-8 TF galaxy distances in each.

\subsection{Cluster Selection}  

The initial characterization of the SCII cluster sample predates this study and originates from preparatory work done by R. Giovanelli for the SCI TF study.  Clusters were selected using the Abell rich cluster catalog as a guide (Abell, Corwin \& Olowin 1980; hereafter ACO).  Expediency played an important role in choosing the parent sample of clusters, of which this sample is a subset.  Nearby clusters (c$z \lesssim$ 10,000 \kms) with preexisting \HI\ velocity widths and $I$ band fluxes were favored, and the more distant clusters (c$z \gtrsim$ 10,000 \kms) that already had a large number of redshifts available were likewise strongly considered as that availability improved the prospects for their kinematical characterization.  Table \ref{tab:SCII} lists the main parameters of the 52 chosen clusters for this work.  
%%%%%%%%%%%%%%%%%%%%%%%%%%%%%%
\begin{table}[!ht]
\caption[The SCII Cluster Sample]{The SCII Cluster Sample}
{\small % get from running rd_to_xyzN.f inside clusters/programs (on done.list)
\def\vcmb{c$z_{\rm cmb}$}
\def\vhel{c$z_{\rm helio}$}
\def\kms2{\small km ${\rm s^{-1}}$}
\def\year{\small B1950}

\begin{center}
\begin{tabular}{lcccrrrcl} \hline \hline
Cluster& R.A.   & Dec.    & $(l,b)$   & \vhel      & \vcmb &$N_z$& R & B-M     \\ 
       & \year  & \year   & \year     & \kms2      & \kms2 &     &   &         \\
\hline
A2806  & 003754 & $-$562600 & (306,$-$61) &  8019(080) &  7867 &  27 & 0 & I-II    \\
A~114  & 005112 & $-$215800 & (128,$-$85) & 17436(143) & 17144 &  41 & 0 & ---     \\
A~119  & 005348 & $-$013200 & (126,$-$64) & 13470(085) & 13141 &  73 & 1 & II-III  \\
A2877  & 010736 & $-$461000 & (293,$-$71) &  7165(058) &  6974 & 130 & 0 & I       \\
A2877b & 010736 & $-$461000 & (293,$-$71) &  9231(048) &  9040 &  37 & - & ---     \\
A~160  & 011012 & $+$151500 & (131,$-$47) & 12390(141) & 12072 &  32 & 0 & III     \\
A~168  & 011236 & $-$000100 & (136,$-$62) & 13365(058) & 13049 &  67 & 2 & II-III: \\
A~194  & 012300 & $-$014600 & (142,$-$63) &  5342(037) &  5037 & 123 & 0 & II      \\
A~260  & 014900 & $+$325500 & (137,$-$28) & 10924(111) & 10664 &  45 & 1 & II      \\
A~397  & 025412 & $+$154500 & (162,$-$37) &  9803(078) &  9594 &  42 & 0 & III     \\
A3193  & 035654 & $-$522900 & (262,$-$47) & 10559(112) & 10522 &  32 & 0 & I       \\
A3266  & 043030 & $-$613500 & (272,$-$40) & 17775(061) & 17782 & 328 & 2 & I-II    \\
A~496  & 043118 & $-$132100 & (210,$-$36) &  9860(059) &  9809 & 148 & 1 & I:      \\
A3381  & 060806 & $-$333500 & (240,$-$23) & 11410(048) & 11510 &  41 & 1 & I       \\
A3407  & 070342 & $-$490000 & (259,$-$18) & 12714(136) & 12861 &  14 & 1 & I       \\
A~569  & 070524 & $+$484200 & (169,$+$23) &  5927(043) &  6011 &  55 & 0 & II:     \\
A~634  & 081030 & $+$581200 & (159,$+$34) &  7822(042) &  7922 &  51 & 0 & III     \\
A~671  & 082524 & $+$303500 & (193,$+$33) & 15092(194) & 15307 &  26 & 0 & II-III: \\
A~754  & 090624 & $-$092600 & (239,$+$25) & 16282(082) & 16599 &  90 & 2 & I-II:   \\
A~779  & 091648 & $+$335900 & (191,$+$44) &  6967(101) &  7211 &  50 & 0 & I-II:   \\
A~957  & 101124 & $-$004000 & (243,$+$43) & 13464(120) & 13819 &  43 & 1 & I-II:   \\
A1139  & 105530 & $+$014600 & (251,$+$53) & 11851(071) & 12216 &  26 & 0 & III     \\
A1177  & 110648 & $+$215800 & (221,$+$66) &  9755(081) & 10079 &   8 & 0 & I       \\
A1213  & 111348 & $+$293200 & (202,$+$69) & 14006(090) & 14304 &  36 & 1 & III     \\
A1228  & 111848 & $+$343600 & (187,$+$69) & 10517(034) & 10794 &  41 & 1 & II-III  \\
A1314  & 113206 & $+$491900 & (152,$+$64) &  9764(154) &  9970 &  28 & 0 & III     \\
A3528  & 125136 & $-$284500 & (304,$+$34) & 16459(139) & 16770 &  30 & 1 & II      \\
A1736  & 132406 & $-$265100 & (313,$+$35) & 10397(050) & 10690 &  62 & 0 & III     \\
A1736b & 132406 & $-$265100 & (313,$+$35) & 13724(084) & 14017 & 116 & - & ---     \\
A3558  & 132506 & $-$311400 & (312,$+$31) & 14342(044) & 14626 & 482 & 4 & I       \\
A3566  & 133606 & $-$351800 & (314,$+$26) & 15370(087) & 15636 &  32 & 2 & II      \\
A3581  & 140436 & $-$264700 & (323,$+$33) &  6865(126) &  7122 &  31 & 0 & I       \\
A1983b & 144724 & $+$170600 & (019,$+$61) & 11332(062) & 11524 &  27 & - & ---     \\
A1983  & 145024 & $+$165700 & (019,$+$60) & 13527(045) & 13715 &  88 & 1 & III:    \\
A2022  & 150212 & $+$283700 & (043,$+$61) & 17262(072) & 17412 &  24 & 1 & III     \\
A2040  & 151018 & $+$073700 & (009,$+$51) & 13440(061) & 13616 &  37 & 1 & III     \\
A2063  & 152036 & $+$084900 & (013,$+$50) & 10445(053) & 10605 & 125 & 1 & II:     \\
A2147  & 160000 & $+$160200 & (029,$+$44) & 10493(085) & 10588 &  93 & 1 & III     \\
A2151  & 160300 & $+$175300 & (032,$+$45) & 11005(059) & 11093 & 143 & 2 & III     \\
A2256  & 170636 & $+$784700 & (111,$+$32) & 17442(132) & 17401 &  94 & 2 & II-III: \\
A2295b & 175900 & $+$691600 & (100,$+$30) & 18701(082) & 18633 &   6 & - & ---     \\
A2295  & 180018 & $+$691300 & (099,$+$30) & 24622(199) & 24554 &   9 & 0 & II-III  \\
A3656  & 195712 & $-$384000 & (002,$-$29) &  5750(064) &  5586 &  31 & 0 & I-II    \\
\hline
\end{tabular}
\end{center}
 }
\label{tab:SCII}
\end{table}
\begin{table}[!ht]
\centerline{Table \ref{tab:SCII} (Continued)}
{\small % get from running rd_to_xyzN.f inside clusters/programs (on done.list)
\def\vcmb{c$z_{\rm cmb}$}
\def\vhel{c$z_{\rm helio}$}
\def\kms2{\small km ${\rm s^{-1}}$}
\def\year{\small B1950}

\begin{center}
\begin{tabular}{lcccrrrcl} \hline \hline
Cluster& R.A.   & Dec.    & $(l,b)$   & \vhel      & \vcmb &$N_z$& R & B-M     \\    
       & \year  & \year   & \year     &  \kms2     & \kms2 &     &   &         \\
\hline
A3667  & 200830 & $-$565800 & (341,$-$33) & 16582(094) & 16477 & 128 & 2 & I-II    \\
A3716  & 204754 & $-$525400 & (346,$-$39) & 13763(064) & 13618 & 174 & 1 & I-II:   \\
A3744  & 210418 & $-$254100 & (021,$-$40) & 11386(089) & 11123 &  42 & 1 & II-III  \\
A2457  & 223312 & $+$011300 & (069,$-$47) & 17643(110) & 17280 &  20 & 1 & I-II:   \\
A2572  & 231554 & $+$182800 & (094,$-$39) & 11857(100) & 11495 &  38 & 0 & III     \\
A2589  & 232130 & $+$163300 & (095,$-$41) & 12288(095) & 11925 &  58 & 0 & I       \\
A2593  & 232200 & $+$142200 & (093,$-$43) & 12414(086) & 12049 &  67 & 0 & II      \\
A2657  & 234218 & $+$085200 & (097,$-$50) & 12028(137) & 11662 &  40 & 1 & III     \\
A4038  & 234506 & $-$282500 & (025,$-$76) &  9012(063) &  8713 & 180 & 2 & III     \\
\hline 
\end{tabular}
\end{center}
 }
\end{table}
%%%%%%%%%%%%%%%%%%%%%%%%%%%%%
The parameters listed include:\newline
Column 1: Standard name according to Abell/ACO catalog.  \newline
Columns 2 and 3: Adopted coordinates of the cluster center, for the epoch 1950; they are obtained from ACO, except for the entries A1983b and A2295b, systems found to be slightly offset from A1983 and A2295 in both sky position and redshift. \newline
Columns 4 and 5: systemic velocities in the heliocentric and in the cosmic microwave background reference frame, respectively, where we assume the motion of the Sun with respect to the CMB is 369.5 \kms\ towards ($l,b$)=(264\fdg4,48\fdg4) (Kogut \etal\ 1993). For all the clusters we derive a new systemic velocity, combining the redshift measurements available in the NED\footnote{The NASA/IPAC Extragalactic
Database is operated by the Jet Propulsion Laboratory, California Institute of 
Technology, under contract with NASA.} database with our own measurements.  An estimated error for the systemic velocity is parenthesized after the heliocentric figure. \newline
Column 6: the number of cluster member redshifts used in determining systemic 
velocities.  \newline
Column 7: Abell richness class. \newline
Column 8: Bautz-Morgan code (Bautz \& Morgan 1970) as listed in the ACO.

\begin{figure}[!ht]
\centerline{\psfig{figure=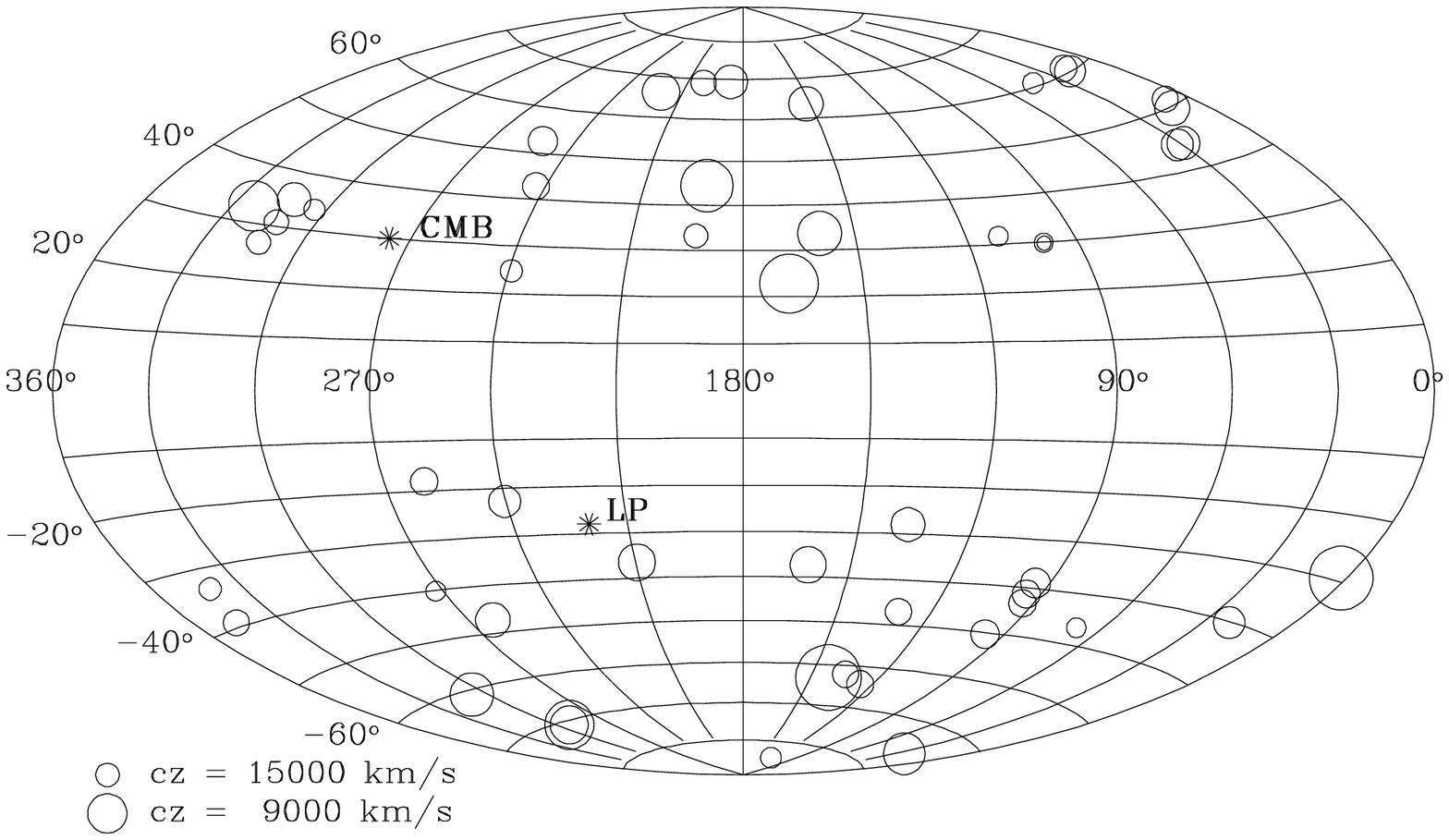,width=5.5in,bbllx=30pt,bblly=234pt,bburx=585pt,bbury=560pt}}
%\centerline{\psfig{figure=figs/aitoff_sample.ps,width=5.0in,angle=270}}
%\vskip -0.5in
\caption[The SCII Sample]
{\ The all-sky distribution of the SCII sample in Galactic coordinates.  The  circles are drawn inversely proportional to the cluster redshifts.  The two examples in the lower left give the scale.  Asterisks mark the apices of the motion of the Local Group with respect to the CMB and the LP cluster inertial frame.}
\label{fig:aitoff_sample}
\end{figure}

\begin{figure}[!ht]
\centerline{\psfig{figure=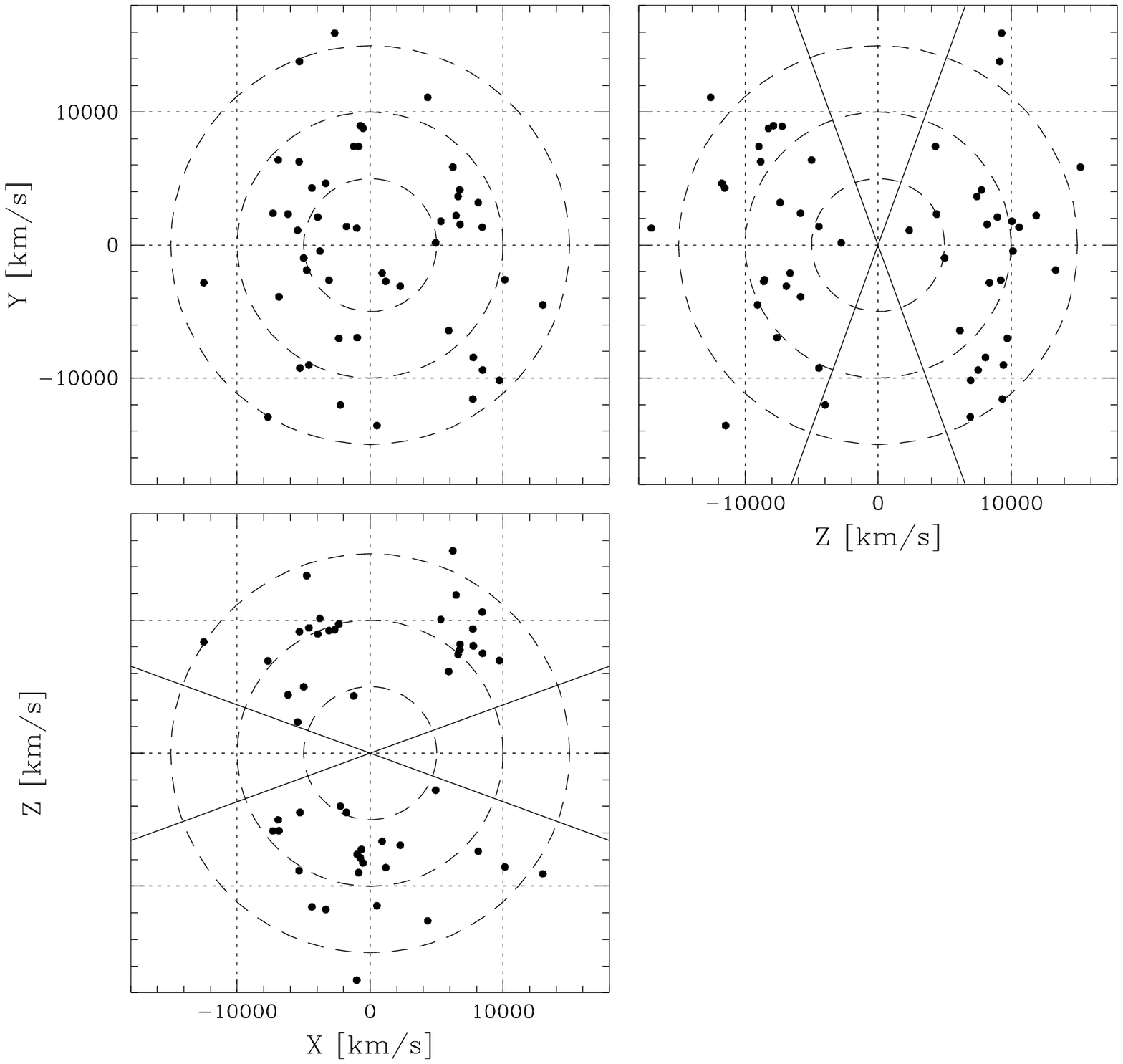,width=5.0in,bbllx=18pt,bblly=150pt,bburx=552pt,bbury=700pt}}
%\vskip -0.5in
\caption[The Sample in Stereo]
{\ The stereographic representation in Galactic Cartesian coordinates of the 52 clusters of galaxies that make up the SCII sample.  The dashed circles have radii of 5000, 10000, and 15000 \kms.  The solid lines in the [X,Z] and [Z,Y] plots indicate $|b| = 20$\deg, the approximate extent of the Zone of Avoidance.}
\label{fig:stereo}
\end{figure}

Figure \ref{fig:aitoff_sample} displays the sample in Galactic coordinates.  The symbol sizes are inversely proportional to the cluster redshifts; two examples are given in the lower left for scale.
%%%%%%%%%%%%%%%%%%%%%%%%%%%%%%%%%%%%%%
%%%%%%%%%%%%%%%%%%%%%%%%%%%%%%%%%%%
An alternative display of the sample is shown in Figure \ref{fig:stereo}, a stereographic view of the sample in Galactic Cartesian 
coordinates.  
%%%%%%%%%%%%%%%%%%%%%%%%%%%%%%%%%%%%%%
%%%%%%%%%%%%%%%%%%%%%%%%%%%%%%%%%%%
The dashed circles have radii of 5, 10, and 15 thousand \kms\ and the solid lines in the [X,Z] and [Z,Y] plots are for $|b| = 20$\deg\ and identify the ZoA.  This distribution recalls our first criterion in selecting the sample: The data set should uniformly sample as much of the sky as prudently feasible.  Since the main thrusts of this work are to recover a bulk flow measurement and to accurately determine a kinematical offset even in the presence of such a flow, an all-sky sample is required.  Unfortunately, the paucity of clusters and the large and uncertain Galactic extinction in the direction of the ZoA prohibit us from sampling that portion of the sky.  In addition, the likelihood of foreground star contamination increases dramatically for objects at low Galactic latitudes.  Our formal criterion was to select clusters at $|b| \gtrsim$ 20\deg\ from the whole sky. 

\begin{figure}[!ht]
\centerline{\psfig{figure=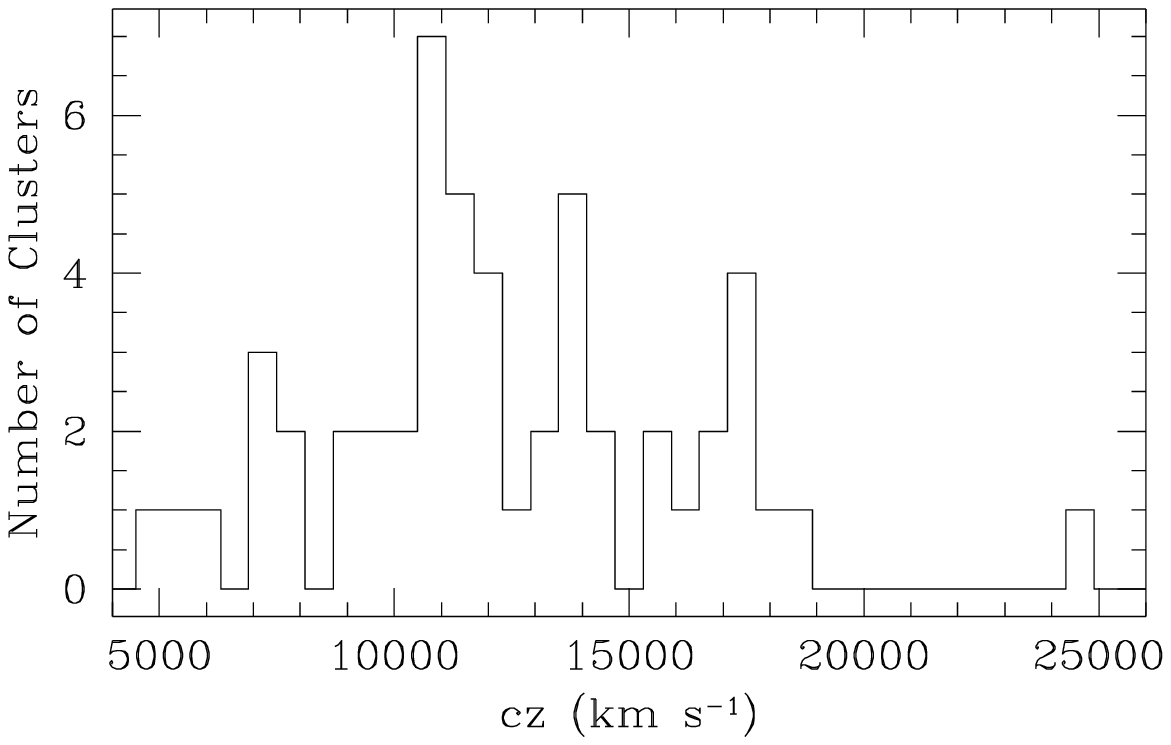,width=4.0in,bbllx=124pt,bblly=279pt,bburx=465pt,bbury=500pt}}
%\vskip -0.5in
\caption[The Cluster Redshift Distribution]
{\ The redshift distribution of the clusters.}
\label{fig:czhist}
\end{figure}

A histogram of the SCII redshift distribution is presented in Figure \ref{fig:czhist}.  The range of redshifts runs from 5000 \kms\ to 25,000 \kms, with fully 90\% of the clusters lying between 7000 \kms\ and 19,000 \kms.  The average CMB redshift of the SCII clusters is 12,050 \kms\ when clusters are weighted according to the square root of the number of TF measurements available.  It is evident from Figure \ref{fig:czhist} that our sample's distance range provides an opportunity to effectively test claims of bulk flow motions on scales of 100\h\ Mpc.
%%%%%%%%%%%%%%%%%%%%%%%%%%%%%%%%%%%%%%
%%%%%%%%%%%%%%%%%%%%%%%%%%%%%%%%%%%

\subsection{Galaxy Selection}

To determine locations of target fields to be imaged, we visually scanned the Palomar Observatory Sky Survey plates for regions in the clusters containing promising disk systems appropriate for TF work.  The selection of target galaxies for this study stemmed directly from the images obtained.  A discussion of the imaging for this project is contained in Papers I, II, and IV.  Each image in clusters chosen for spectroscopy was searched for spiral disks with the following properties:  \newline
(i) disk inclinations $\gtrsim$ 40$^{\circ}$; \newline
(ii) lack of dominating bulges -- bulgy disk systems tend to be gas deficient and thus undesirable for emission line spectroscopy.  Moreover, morphological homogeneity is preferable, to limit the effects of morphological bias (Section \ref{sec:morph}); \newline
(iii) no apparent warps/interacting neighbors; and \newline
(iv) no nearby bright stars which may affect flux measurements.  

It should be noted that the above properties served as guidelines for selecting TF candidates, but occasionally the guidelines were not strictly followed, as occasionally our hand was forced by the vagaries of telescope allocations and weather conditions.  Coordinates and position angles for approximately 2250 TF candidates were obtained from the Digitized Sky Survey\footnote{The Digitized Sky Surveys were produced at the Space Telescope Institute under U.S. Government grant NAG W-2166.  The images of these surveys are based on photographic data obtained using the Oschin Schmidt Telescope on Palomar Mountain and the UK Schmidt Telescope.} and are accurate to better than 2\arcsec.  The full list is available from D.A.D. upon request.

\begin{figure}[!ht]
\centerline{\psfig{figure=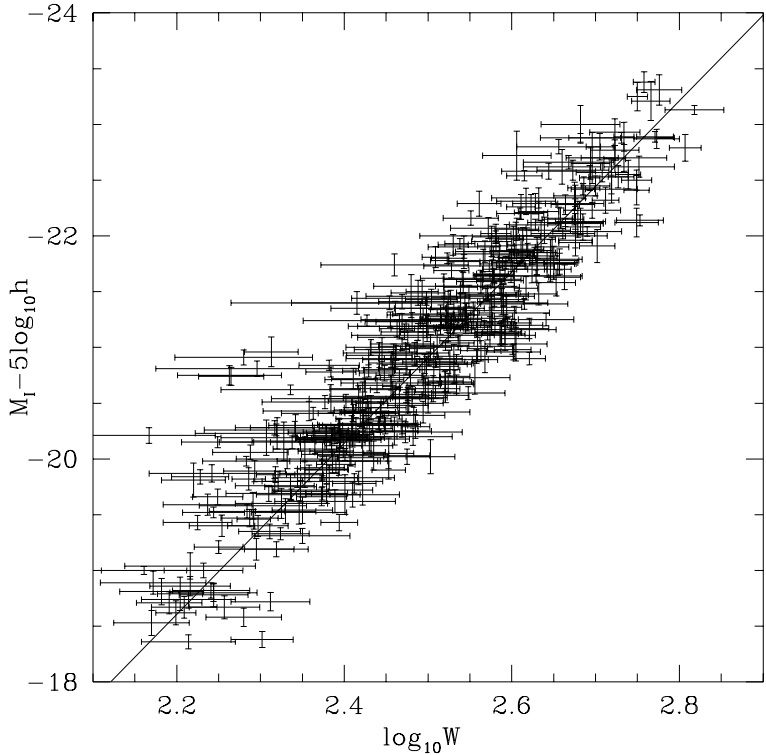,width=4in,bbllx=15pt,bblly=150pt,bburx=566pt,bbury=696pt}}
\caption[Tully-Fisher Data]
{\ The TF data for all galaxies is displayed above.  We emphasize that the data have {\it not} yet been corrected for individual cluster peculiar motion and incompleteness bias.  The solid line is the template relation derived later in Section \ref{sec:results}.}
\label{fig:TFall_uncor}
\end{figure}

A TF display of the collective raw data can be seen in Figure \ref{fig:TFall_uncor}.  The term ``raw'' implies that the absolute magnitudes $M_I$ and the rotational velocity widths $W$ are corrected for all the effects described in Papers I and II (extinction, inclination, etc), but the effects of cluster incompleteness bias and peculiar motion are not accounted for.  Included is the template relation obtained later in Section \ref{sec:results} (cf. Equation \ref{eq:TF}).  Hereafter we will use $y=M_I-5\log h$ and $x=\log W-2.5$.
%%%%%%%%%%%%%%%%%%%%%%%%%%%%%%%%%%%%%%
%%%%%%%%%%%%%%%%%%%%%%%%%%%%%%%%%%%

%\section{THE TULLY-FISHER RELATION}
\section{The Tully-Fisher Relation}

The most widely applied methods for the determination of redshift-independent 
distances of galaxies rely on the combination of photometric (distance 
dependent) and kinematic (distance independent) parameters.  Among them, and 
arguably the most accurate, are the TF and fundamental plane relations, used for spiral and elliptical galaxies, respectively.  With both these methods distances for individual galaxies, and therefore peculiar velocities, are computed by comparison with a fiducial {\it template} relation which must be observationally derived.  The template relation defines the rest reference frame, against which peculiar velocities are to be measured.  The importance of accurately calibrating such a tool cannot be overemphasized.  It is possible that the discrepancies between different claims of bulk motions may be partly related to insufficiently well determined template relations.  

Ideally, the template employed is tied to a kinematical rest frame.  In practice, the template originates from data within the sampled volume itself.  We outline here our approach to calibrate the TF relation using our sample of data from 52 rich Abell clusters distributed throughout the sky.  As we shall see, the TF zero point is obtained through an iterative process: the computation of the cluster incompleteness bias requires prior knowledge of the TF template zero-point, while the computation of the zero point demands a correction for the cluster incompleteness bias.  The process does converge, however, in a small number of iterations.

\subsection{The Calibration of the Tully-Fisher Relation}
\label{sec:calibration}

For an assumed linear TF relation we need to determine two main parameters: a slope and a magnitude offset, or zero point.  The slope of the TF relation is best determined by a sample that maximizes the dynamic range in $(M, \log W)$, i.e. one that preferentially includes nearby objects.  On the other hand, the magnitude zero point of the relation is best obtained from a sample of distant objects for which a peculiar motion of given amplitude translates into a small magnitude shift.  An often-adopted practice to calibrate the TF relation uses one relatively distant cluster of galaxies, for example the rich Coma cluster at c$z$\about 7200 \kms.  Such a choice relies on the assumption that galaxies within a cluster are essentially at the same redshift -- any differences in their radial velocities are attributed to the cluster's virial stretch.  Thus, they all equally participate in the local peculiar velocity field and they should all obey the same local TF relation.  Another reason to choose a cluster like Coma for calibration is that, as mentioned above, the selection of a distant object limits the impact of the object's peculiar velocity $V$, at least to the extent that the physical size of the cluster is small in comparison to its mean distance; peculiar velocities introduce relative distortions of redshifts that are larger for nearby objects than for distant ones.  In terms of magnitude, if $V_{\rm pec}$$<<$c$z$, the zero point will be off by $\Delta m \simeq -2.17V_{\rm pec}/{\rm c}z$ magnitudes (cf. Equation \ref{eq:V}).

There are several problems with the above calibration scheme.  First, a ``template'' cluster needs to have a large sample of spiral galaxies.  Second, even for a cluster as distant as Coma, typical cosmic velocities may bias the relation's zero point.  If the cluster were moving at a plausible speed of 500 \kms, then all other estimates of peculiar velocities that use the template will systematically be off by 500(c$z$/c$z_{\small \rm template}$), regardless of statistical uncertainties.

To avoid such systematics in the TF relation, G97a,b use a ``basket of clusters'' approach.  Their template is derived through an iterative procedure that simultaneously determines the TF zero point and the cluster motions of their sample.  They assume that the mean peculiar velocity of the clusters farther than c$z=4000$ \kms\ is null, i.e. they zero the monopole of the more distant cluster peculiar velocity distribution function.  This approach does not, however, affect the value of the dipole or higher moments, and thus still allows an effective measure of possible bulk flows.  The proximity of the SCI sample provides the stretch in galactic properties necessary to determine accurately the TF slope (it is for this reason that we adopt the slope computed in G97b).  Also, the large database afforded by the SCI allows the statistical uncertainty of the magnitude zero point to be reduced to $0.02-0.03$ magnitudes.  The systematic uncertainty in the zero point is larger.  For a sample of $N$ objects with an rms velocity of $\left<V^2\right>^{1/2}$ at a mean redshift of $\left<{\rm c}z\right>$, the expected accuracy of the zero point is limited by systematic concerns to
\be
\sigma_a \approx {2.17 \left<V^2\right>^{1/2} \over \left<{\rm c}z\right> \sqrt{N}} \;\;\; {\rm mag}.
\label {eq:sigma_a}
\ee
G97b conclude the SCI systematics only allows the universal TF zero point to be determined to within 0.04 magnitudes.  Thus the overall zero point accuracy obtainable from the SCI sample is \about 0.05 magnitudes.  We adopt the procedure described in G97b to calibrate the TF zero point using the SCII sample.  The resulting improvement in its calibration is discussed in Section \ref{sec:results}.

\subsection{The Scatter of the Tully-Fisher Relation}
\label{sec:scatter}

Any relation involving observed parameters has a limited accuracy described by the amplitude of the relation's scatter.  Claims of the scatter in the TF relation vary from as low as 0.10 mag (Bernstein \etal\ 1994) to as high as 0.7 mag (Sandage \etal\ 1994a,b; 1995; Marinoni \etal\ 1998).  The amplitude of the scatter does depend on wavelength, and studies in the $I$ band typically yield the tightest relations.  The efforts with the largest samples yield 1$\sigma$ dispersion values of \about 0.3--0.4 mag (Mathewson, Ford \& Buckhorn 1994; Willick \etal\ 1995; G97b).  Uncertainties in observational measurements are not the only factors that lead to the overall spread in the data.  The corrections to the observed fluxes and disk rotational velocities described in Papers I and II are not exactly known, nor are the methods we detail later to account for inherent sample biases such as cluster incompleteness.  Moreover, there is an {\it intrinsic} component to the TF dispersion since individual galaxies have diverse formation histories.  In fact, Eisenstein \& Loeb (1996) advocate an intrinsic scatter of 0.3 magnitudes, a number greater than most estimates from observational work.  In light of this fact, they make the interesting claim that either spirals formed quite early or that there must be a type of feedback loop that promotes galactic assimilation.

\begin{figure}[ht]
\centerline{\psfig{figure=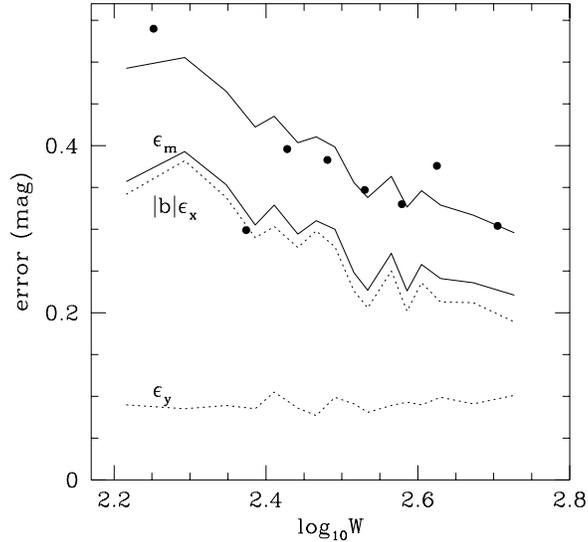,width=3.0in,bbllx=52pt,bblly=185pt,bburx=473pt,bbury=595pt}}
\caption[The Scatter in the Tully-Fisher Relation]
{\ The observed scatter in the TF relation.  Running averages (in units of magnitudes) of the observed errors are given as dotted lines.  The lower solid line is the sum in quadrature of the two dotted lines and represents the total measurement uncertainty.  The circles plotted indicate the standard deviations of residuals from the template TF relation.  The upper solid line is the total scatter in the TF relation and has been computed by a sum in quadrature of the observed error distribution and an intrinsic component equivalent to that found in G97b (see Equation \ref{eq:int} below).}
\label{fig:scatter}
\end{figure}

As already established in G97b and Willick (1999), Figure \ref{fig:scatter} reinforces the notion of low intrinsic scatter.  
%%%%%%%%%%%%%%%%%%%%%%%%%%%%%%%%%%%%%%
%%%%%%%%%%%%%%%%%%%%%%%%%%%%%%%%%%%
The two dotted lines indicate the velocity width and magnitude uncertainties $\epsilon_x$ and $\epsilon_y$, with $\epsilon_x$ multiplied by the TF slope $b$ to put it on a magnitude scale; the solid line labeled $\epsilon_m=\sqrt{(b \epsilon_x)^2 + \epsilon_y^2}$ represents the average measurement uncertainty.  The data displayed in Figure \ref{fig:scatter} are generated using equal numbers of galaxies per data point.  The circles plotted represent the average standard deviations of the residuals from the fiducial TF relation.  We see that the velocity width errors dominate those from the $I$ band fluxes, which are approximately independent of velocity width.  Furthermore, the logarithmic velocity widths become increasingly uncertain for slower rotators (cf. G97b; Willick \etal\ 1997).  We approximate the total observed scatter with a simple linear relation that depends on the velocity width:
\be
\sigma_{\rm tot} = -0.40x + 0.38 \; {\rm mag}.
\label{eq:scat}
\ee

The total scatter found here is in general larger than that found in G97b ($\sigma_{\rm tot} = -0.32x + 0.32$).  This is unsurprising given our use of optical rotation curves instead of 21 cm profiles, a comparatively easier source from which to estimate velocity widths, and since the nearer SCI galaxies generally have better determined disk inclinations.  The gap between the observed scatter and the measured errors is attributed to an intrinsic scatter contribution: the top line is a sum in quadrature of our observed measurement errors, $\epsilon_m$, and the intrinsic scatter found in G97b:
\be
\sigma_{\rm int} = -0.28x + 0.26 \; {\rm mag}.
\label{eq:int}
\ee
Simply put, our findings for an intrinsic component of the TF relation agree with those of Giovanelli and Willick and their collaborators.

\subsection{Observational Biases}
\label{sec:bias}

\subsubsection{Cluster Population Incompleteness}
\label{sec:CI}

Quite possibly the most important selection effect to consider for this observing program is that of cluster incompleteness, the preferential sampling of the bright end of the luminosity function (LF).  Several authors have cautioned that cluster incompleteness can significantly alter inferred distances to clusters, though this ultimately depends on the amplitude of the scatter in the TF relation (see G97b, Willick 1994, Sandage, Tammann \& Federspiel 1995, and the review by Teerikorpi 1997).  Schechter (1980) first advanced the notion that the repercussions of this selection effect can be circumvented through an inverse fitting procedure, whereby the roles of the ordinate and abscissa in the TF diagram are reversed.  The argument goes as follows: if there are no observational effects working against the selection of velocity widths and if the errors on the absolute magnitudes are negligible, then a fit to log$W$ vs. $M$ will not be affected by a cutoff at faint magnitudes.  Unfortunately, magnitude errors cannot be ignored (see Figure \ref{fig:scatter}), and moreover, the errors in $M$ and log$W$ are coupled through inclination corrections; TF data do not obey a sharp faint magnitude limit.  Inevitably, the absence of faint galaxies in a TF sample artificially brightens the zero point and lowers the slope.  We describe next our methodology to account for the effect.

Our first concern is to quantify the characteristics of the observed luminosity distribution with respect to the actual LF.  As an aside, we note that current work by Andreon (1998) supports the notion of a canonical spiral LF by proposing that the LF for the separate morphological classes E, S0, and S is independent of environment -- observed differences in the overall LF for different environments are merely due to varying proportions of the morphological classes (see, however, Iovino \etal\ 1999).  Our first step is to sum up the observed luminosities in bins of absolute magnitude.  Computing such absolute magnitude histograms for each individual cluster, where the membership counts can be as small as \about 5, would not be statistically indicative on the whole.  We therefore compute {\it average} histograms for several distance ranges of width 2000 \kms.
%%%%%%%%%%%%%%%%%%%%%%%%%%%%%%%%%%%%%%
\begin{figure}[ht]
\centerline{\psfig{figure=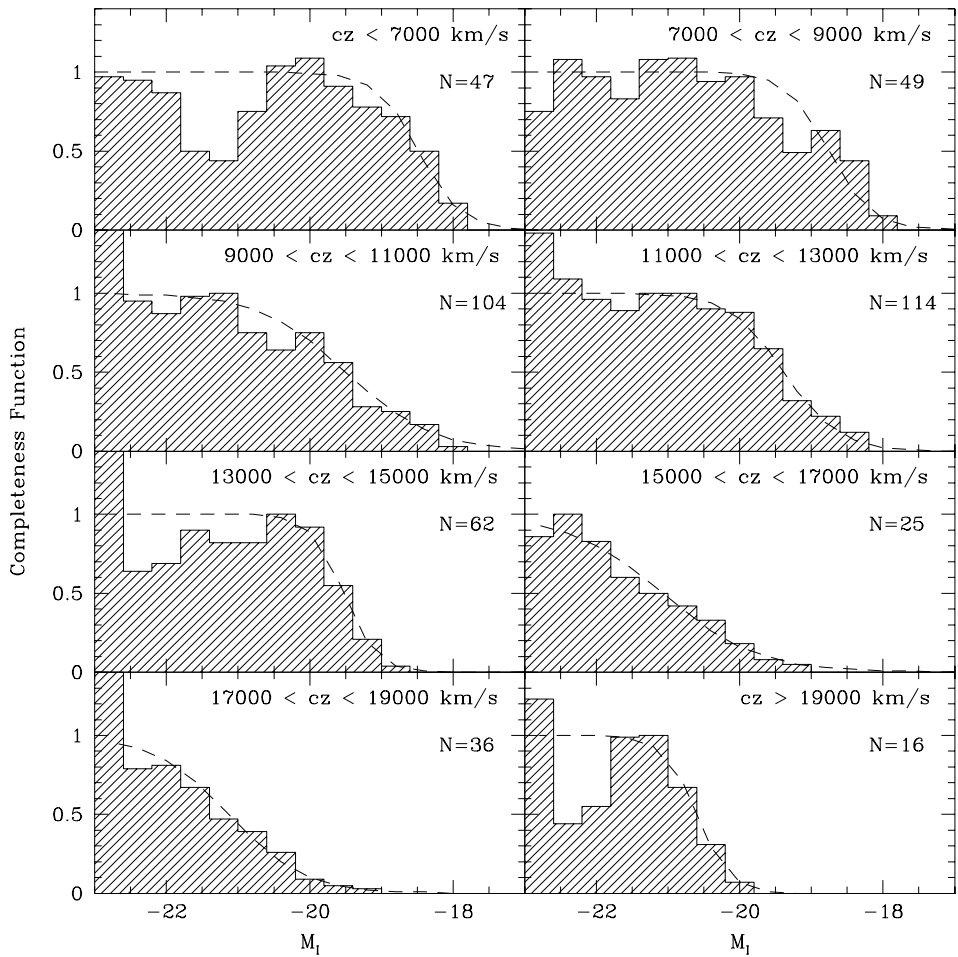,width=5.0in,bbllx=39pt,bblly=159pt,bburx=566pt,bbury=690pt}}
\caption[Completeness Distributions]
{\ The distributions of cluster completeness for separate distance ranges.  The dashed line is our smoothed approximation to the completeness function.  The number of galaxies within each redshift range are indicated.}
\label{fig:complete}
\end{figure}
%%%%%%%%%%%%%%%%%%%%%%%%%%%%%%%%%%%
We define the completeness function as the ratio of the observed luminosity distribution to the assumed intrinsic luminosity distribution.  The following figures have been constructed assuming a Schechter LF with $M^*=-21.6$ and $\alpha=-0.5$ (see Figure 20 of Paper IV), but final TF templates using other Schechter parameters will also be provided.  Figure \ref{fig:complete} displays a smoothed representation of the completeness functions, along with our fitted approximation.  As in G97b, we borrow a relation from Fermi-Dirac statistics to model completeness:
\be
c(y) = {1 \over e^{(y-y_{\rm t})/\beta}+1}
\ee
where $y_{\rm t}$ represents a transition luminosity in the fit, and $\beta$ characterizes the steepness of the drop.  It should be noted that the final TF template is rather robust in terms of the completeness function construction.  The template is largely insensitive to the choice of distance regimes, variations in the Fermi-Dirac fit profile, and the luminosity bin widths.  Reasonable alterations in these parameters induce negligible changes in the TF template zero point.  With the estimation of the completeness function in hand, we can now quantify a given cluster's incompleteness bias via Monte Carlo simulations.

\subsubsection{Monte Carlo Simulations of Cluster Parent Samples}

Our task is simplified by assuming the bias-corrected slope derived in G97b is valid for our sample, a justifiable assumption given the relative propinquity of their data -- a broader dynamic range can be more easily sampled in nearby clusters.  Furthermore, least-square fits to the data yield slopes that, within the errors, agree with the TF slope from G97b.  On a more qualitative level, inspection of Figure \ref{fig:TFall_uncor} indicates our data agree with the slope from G97b.  It is thus sufficient to only determine the bias in the TF offset for each cluster.  

\begin{figure}[ht]
\centerline{\psfig{figure=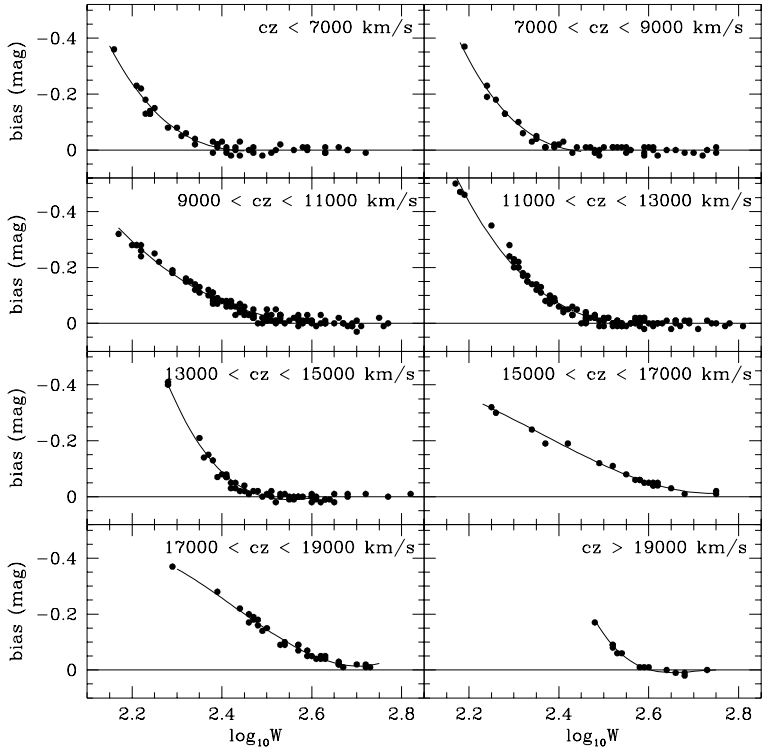,width=4.0in,bbllx=28pt,bblly=159pt,bburx=566pt,bbury=690pt}}
\caption[Incompleteness Bias]
{\ The incompleteness bias as a function of rotational velocity width is shown.
Individual points are the biases generated for each galaxy from Monte Carlo simulations and the solid lines are polynomial fits to the bias.}
\label{fig:bias}
\end{figure}

We compute the bias as follows: for each galaxy a large number 
($N_{\rm iter}=10^3$) of trial TF data points are generated that mimic the general characteristics of the actual data.  Within each trial, a random (Gaussian deviate) magnitude offset $\Delta y_{\rm trial}$ is first chosen according to the TF scatter relation, Equation \ref{eq:scat}, which depends on the galaxy's measured velocity width parameter $x$.  The addition of this offset to the magnitude inferred from the final, template TF relation yields the trial absolute magnitude:
\be
y_{\rm trial}=b_{\rm tf}x+a_{\rm tf}+\Delta y_{\rm trial}.
\ee
The trial magnitude is kept only if it is a likely magnitude, i.e. a second random number from the interval [0,1] must be less than $c(y_{\rm trial})$, the completeness value for the trial luminosity.  Otherwise, the process is repeated until a likely magnitude is found.  After all $N_{\rm iter}$ iterations are complete for a given galaxy, the incompleteness bias is taken to be the mean difference between the trial luminosities and that expected from the TF relation, i.e. 
\be
-\Delta y_{\rm icb}={\sum_i^{N_{\rm iter}} y_{{\rm trial},i} \over N_{\rm iter}} -(b_{\rm tf}x+a_{\rm tf}).
\ee
Figure \ref{fig:bias} gives the biases calculated as a function of velocity width and the polynomials we fit to them.  
%%%%%%%%%%%%%%%%%%%%%%%%%%%%%%%%%%%%%%
%%%%%%%%%%%%%%%%%%%%%%%%%%%%%%%%%%%
The general property of the computed bias is as expected: incompleteness biases ``turn on'' at higher velocity widths for more distant objects.  An incompleteness bias-corrected zero point for each cluster is then extracted from the cluster's distribution of observed velocity widths and bias-corrected absolute magnitudes.  A tabulation of the offsets from the template zero point,  $a-a_{\rm tf}$, is provided in Table \ref{tab:offsets}.  The cluster names are listed first and are followed by the number of cluster members $N_{\rm tf}$ with reliable photometry and velocity widths.  The next two columns are measures of the cluster offsets from the template zero point, the second of which is corrected for the effect of cluster incompleteness and includes an indication of its uncertainty $\epsilon_a$ in parentheses, e.g. $-0.06$(10) implies $-0.06 \pm 0.10$ magnitudes.  The last column of data listed is $\sigma_a$, the dispersion in the difference between the template and the cluster's set of absolute magnitudes.  It is used to compute the offset uncertainty: 
\be
\epsilon_a = \sigma_{\rm max}/\sqrt{N_{\rm tf}},
\label{eq:epsilon_a}
\ee
where $\sigma_{\rm max}$ is taken to be the maximum of [0.35 mag, $\sigma_a$] to avoid overly optimistic measures of offset uncertainty in cases of chance alignment of the data due to small number statistics.  Lastly, we remark that the proper incorporation of selection effect corrections is vital to determining the TF template, but the exact evaluation of such biases necessarily requires a TF template.  A stable solution to this circular process is found within a few iterations.
%%%%%%%%%%%%%%%%%%%%%%%%%%%%%%
\begin{table}
\caption[Cluster Offsets]{Cluster Offsets}
{\small \def\ea{$\epsilon_a$}
\def\sa{$\sigma_a$}
\def\ab{$a_{\rm bias}-a_{\rm tf}$}
\def\au{$a-a_{\rm tf}$}
\def\N{$N_{\rm tf}$}

\begin{center}
\begin{tabular}{|lrccc|lrccc|} \hline \hline
Cluster & \N  &  \ab   &   \au     &  \sa & Cluster & $N$ & \ab    &   \au     &  \sa \\
\hline
A2806   & 10  & -0.15 & -0.13(11) & 0.27 & A3528   &  3  & ~0.03 & ~0.18(07) & 0.12 \\
A 114   &  9  & -0.01 & ~0.07(14) & 0.41 & A1736   &  6  & -0.03 & ~0.01(18) & 0.44 \\
A 119   &  6  & ~0.04 & ~0.05(16) & 0.38 & A1736b  &  4  & -0.03 & -0.03(18) & 0.32 \\
A2877   &  7  & -0.01 & ~0.03(15) & 0.40 & A3558   &  8  & -0.13 & -0.10(15) & 0.43 \\
A2877b  &  5  & -0.10 & -0.08(16) & 0.14 & A3566   &  9  & -0.13 & -0.03(12) & 0.23 \\
A 160   &  6  & -0.06 & -0.05(18) & 0.45 & A3581   &  4  & -0.01 & ~0.04(20) & 0.40 \\
A 168   &  9  & -0.13 & -0.12(13) & 0.38 & A1983b  &  8  & -0.27 & -0.26(12) & 0.25 \\
A 194   & 13  & ~0.05 & ~0.09(13) & 0.45 & A1983   &  7  & -0.14 & -0.07(19) & 0.50 \\
A 260   &  9  & ~0.13 & ~0.23(14) & 0.42 & A2022   &  8  & ~0.06 & ~0.14(12) & 0.32 \\
A 397   & 14  & -0.17 & -0.13(15) & 0.57 & A2040   & 10  & -0.10 & -0.03(14) & 0.43 \\
A3193   &  6  & -0.14 & -0.10(14) & 0.09 & A2063   & 18  & -0.18 & -0.14(09) & 0.36 \\
A3266   &  2  & ~0.12 & ~0.31(25) & 0.21 & A2147   & 19  & -0.09 & -0.06(09) & 0.39 \\
A 496   &  9  & -0.18 & -0.13(12) & 0.36 & A2151   & 22  & -0.08 & -0.06(08) & 0.40 \\
A3381   &  4  & -0.17 & -0.16(18) & 0.12 & A2256   &  8  & -0.04 & -0.01(12) & 0.33 \\
A3407   &  8  & -0.07 & ~0.03(21) & 0.58 & A2295b  &  4  & -0.05 & ~0.05(18) & 0.36 \\
A 569   & 13  & ~0.02 & ~0.06(10) & 0.30 & A2295   & 10  & ~0.07 & ~0.10(12) & 0.39 \\
A 634   &  8  & ~0.04 & ~0.06(12) & 0.35 & A3656   &  6  & ~0.01 & ~0.03(14) & 0.28 \\
A 671   &  9  & -0.05 & ~0.02(12) & 0.34 & A3667   &  4  & ~0.27 & ~0.37(18) & 0.28 \\
A 754   &  3  & -0.07 & ~0.01(42) & 0.73 & A3716   & 14  & -0.08 & -0.06(09) & 0.22 \\
A 779   & 14  & ~0.01 & ~0.03(09) & 0.27 & A3744   & 11  & -0.03 & ~0.03(11) & 0.37 \\
A 957   &  6  & ~0.08 & ~0.13(14) & 0.16 & A2457   &  9  & -0.02 & ~0.02(12) & 0.32 \\
A1139   & 11  & -0.17 & -0.13(12) & 0.39 & A2572   &  5  & -0.11 & -0.08(16) & 0.13 \\
A1177   &  6  & -0.11 & -0.01(15) & 0.37 & A2589   &  6  & -0.03 &  0.04(14) & 0.23 \\
A1213   &  6  & -0.12 & -0.12(14) & 0.33 & A2593   & 12  &  0.09 &  0.13(10) & 0.30 \\
A1228   & 13  &  0.04 &  0.12(10) & 0.33 & A2657   &  5  & -0.03 & -0.01(16) & 0.26 \\
A1314   &  8  & -0.04 &  0.03(12) & 0.31 & A4038   &  7  & -0.06 & -0.02(13) & 0.35 \\
\hline
\end{tabular}
\end{center}
 }
\label{tab:offsets}
\end{table}
%%%%%%%%%%%%%%%%%%%%%%%%%%%%%

\subsubsection{Homogeneous Malmquist Bias}

An observational bias commonly referred to as the ``homogeneous Malmquist bias'' is a distance underestimate for a galaxy drawn from a uniform distribution of galaxies.  The bias is a direct result of the error on the measured distance -- a galaxy taken from a Poissonian spatial distribution and measured to be at a distance modulus $\mu_{\rm m} \pm \Delta \mu$ is more likely to actually lie between $\mu_{\rm m}+\Delta \mu$ than in the range $\mu_{\rm m}-\Delta \mu$, due to the larger volume of the more distant shell.  Following the reasoning given in Lynden-Bell \etal\ (1988), it can be shown that a measured distance $R_{\rm m}$ is, on average, an underrepresentation of the true distance by the factor $\exp(3.5\Delta^2)$:
\be
%R_\circ=R_{\rm m} e^{3.5\Delta^2}.
V={\rm c}z-H_\circ R_{\rm m} \rightarrow
{\rm c}z-H_\circ R_{\rm m} e^{3.5\Delta^2}.
\ee
The factor $\Delta=10^{0.2\epsilon_a}-1$ is the fractional error in the TF distance measurement, roughly 17\% for individual galaxies and 6\% for our clusters.  It was shown in G97b that such a correction for clusters of galaxies is rather small, almost to the point of being negligible (of order 1\% on individual cluster distances).  The inclusion of the above correction factor in later computations in this work does not change results appreciably.  The influence of the homogeneous Malmquist bias on the inferred local bulk flow is discussed in Paper III.

The Abell/ACO clusters do not represent a homogeneously distributed population.  Thus, the application of a homogeneous Malmquist bias correction may appear incorrect.  However, a correction that takes into account the clustering properties of clusters differs from the homogeneous Malmquist bias correction only in the second order.  Given the miniscule size of the correction and the uncertainty of a possible ``inhomogeneous'' Malmquist bias correction, we consider the application of a homogeneous Malmquist bias correction a satisfactory approach.

\subsubsection{Cluster Size} % used /bias/fat_cluster.f

In calculating absolute magnitudes, we assume that each cluster member within one Abell radius of the cluster center is at a distance equal to the average of all the redshifts available for the cluster from the literature.  It is unlikely that a given galaxy will actually be at this distance, but it is a useful approximation if the distance to the cluster is significantly larger than the cluster's virial size.  For all but a handful of our clusters this is the case, so cluster size biases play a small role in our analysis.

Our concern here is the method of averaging used.  The calculation of a cluster offset involves averaging over absolute magnitudes.  Thus, averages of the logarithms of the distances are computed, when the logarithm of the average distance is actually desired.  This results in a systematic underestimate of cluster distances.  We can use either the angular distribution on the sky of the cluster galaxies or the Abell radius to infer the approximate physical size of the each cluster.  Simple analytic calculations using either estimate of the cluster size show that the amplitude of the bias is at most of order 0.001 magnitudes for even our closest clusters, so we shall not concern ourselves with this effect any further.  A more serious concern, the morphological dependence on the TF relation, is investigated next.  

\subsubsection{Morphology}
\label{sec:morph}

Ideally, scientific models are simple and are thus constructed with the fewest number of free parameters the data demand.  In TF work, it would be preferable to limit the range of morphologies sampled, since physical parameters are known to vary along the Hubble sequence (Roberts \& Haynes 1994).  For example, it has been shown that, for a given (optical) luminosity distribution, the velocity width distribution for early-type galaxies is shifted to higher rotational speeds with respect to the distribution for later types (Roberts 1978).  There is evidence, however, that the TF differences between morphological types appear to diminish at wavelengths longer than $I$ band (e.g. Aaronson \& Mould 1983).  Regardless, we are fortunate that one sample selection criterion of ours, that of disks being rich in ionized gas, encouraged a rather homogeneous sample comprised mainly of Sc types.  Our sample has the following population properties: 14\% are of type earlier than Sb, 16\% are type Sb, and 70\% are classified as Sbc or later.  The majority (52\%) of the galaxies are type Sc.  

\begin{figure}[ht]
%\centerline{\psfig{figure=figs/TFmorph.ps,width=6.0in,bbllx=46pt,bblly=419pt,bburx=572pt,bbury=697pt}}
\centerline{\psfig{figure=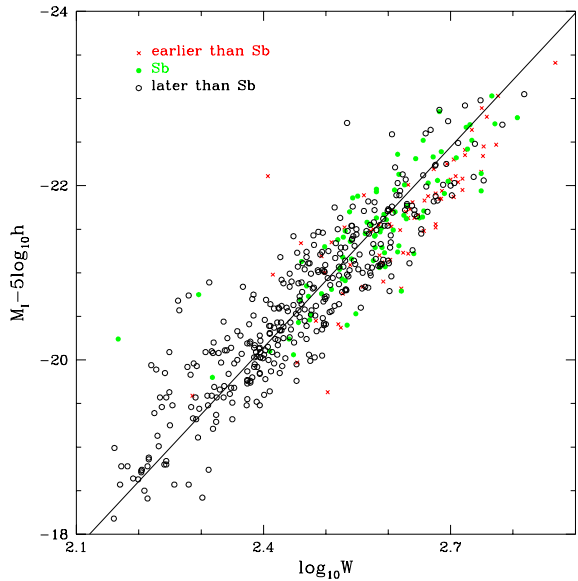,width=3.0in,bbllx=20pt,bblly=140pt,bburx=565pt,bbury=694pt}}
\caption[Morphological Bias]
{\ The TF diagram differentiated according to morphological class.  Asterisks represent galaxies earlier than Sb, filled circles symbolize type Sb, and open circles are for Sbc and later.  The data are corrected for cluster peculiar motions.}
\label{fig:TFmorph}
\end{figure}

Our sample affords statistically significant tests of TF morphological dependencies.  We plot in Figure \ref{fig:TFmorph} the TF parameters with symbols differentiated according to morphological class.  
%%%%%%%%%%%%%%%%%%%%%%%%%%%%%%%%%%%%%%
%%%%%%%%%%%%%%%%%%%%%%%%%%%%%%%%%%%
Filled circles symbolize Sb types, whereas open circles and asterisks represent types later and earlier than Sb, respectively.  The plot includes the data for all galaxies deemed to have reliable velocity widths, with each cluster member corrected for cluster peculiar motion.  The solid line drawn uses the fiducial TF slope of $b_{\rm tf}=-7.68$ and zero point $a_{\rm tf}=-20.91$ mag (cf. Section \ref{sec:results}).  As found in G97b, a clear distinction is evident between the three Hubble types in the form of a fainter zero point for earlier types.  The error-weighted averages of the offsets $\Delta m_T$ from the template zero point for the three morphological classes differ in the following ways for our sample:

{\it Types earlier than Sb:} $-0.27$ $(-0.32)$ mag

{\it Type Sb:} $-0.11$ $(-0.10)$ mag

{\it Other types:} unchanged

\noindent The numbers given in parentheses are those from G97b.  For consistency, we will continue to utilize the offsets obtained in G97b.

\subsubsection{Environment}

Yet another possible bias in our sample is the effect of environment.  For instance, the more distant clusters in our sample (and the Abell/ACO catalog in general) tend to be richer.  A richer cluster typically has a stronger intracluster medium X-ray flux and a more regular, elliptical-dominated core (see, for example, Sarazin 1986).  Spiral galaxies predominantly lie in a rich cluster's periphery and the closer a spiral disk is to the cluster center, the less likely it is to contain neutral hydrogen gas (Giovanelli \& Haynes 1985).  This lack of interstellar gas within clusters of galaxies may be due to evaporation into the hotter intracluster gas, or it may be attributed to ``stripping'' originating from either tidal galaxy--galaxy interactions or ram pressure ablation on intracluster gas.  Rubin, Ford, \& Whitmore (1988) and Whitmore, Forbes, \& Rubin (1988; WFR hereafter) claim that the inner spiral galaxies within clusters exhibit falling rotation curves (RCs), as opposed to the asymptotically flat or rising RCs usually seen in the cluster periphery and field.  Furthermore, they find cluster RCs may be of lower amplitude than field RCs.  They offer the explanation that the falling (and lower amplitude) RCs arise from mass loss ---  the inner galaxies have had their dark matter halos stripped --- or that the cluster environment simply inhibits halo formation.  A related finding by WFR is a monotonic increase in the mass to light ratio, with cluster radius, which they ascribe to the changing RC shape with cluster radius.  This view has been contested, however, by Amram \etal\ (1993) and Vogt (1995) who find little evidence for any gradients in the outer portions of RCs.  If RCs in the inner regions of rich clusters do indeed differ from RCs in other environments, the ramifications are significant.  The dependence of the TF relation and/or its dispersion on environment are possible consequences.  Finally, we point out that rich clusters represent high density peaks and may thus be home to a recent merger; at least one third of all rich clusters are home to a recent inhomogeneous superposition of two or more separate systems (Girardi \etal\ 1997).  It is prudent to verify whether environment plays a significant role in our data.

\begin{figure}[ht]
\centerline{\psfig{figure=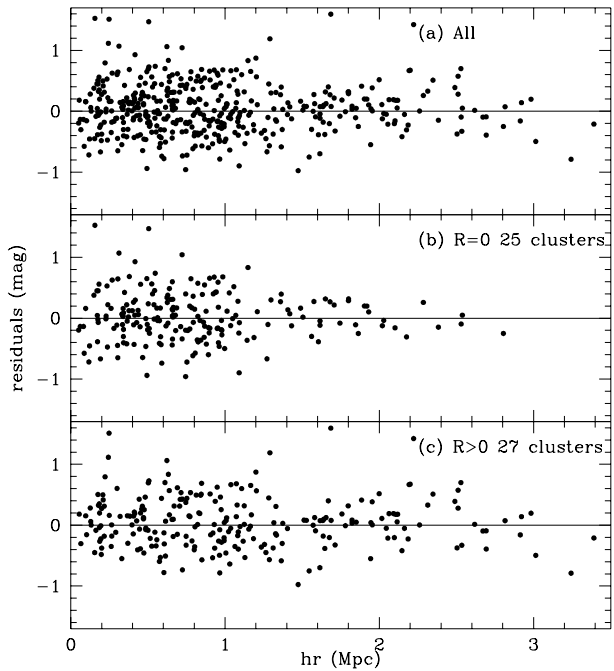,width=3.0in,bbllx=62pt,bblly=143pt,bburx=547pt,bbury=707pt}}
\caption[Environmental Bias]
{\ A test of environmental bias is displayed.  TF residuals are given versus
projected physical distance from the cluster center.  The data are corrected for peculiar velocity and morphological offsets.  Panel (a) displays all the cluster data, whereas panels (b) and (c) differentiate between cluster richness.}
\label{fig:env}
\end{figure}

Figure \ref{fig:env} displays one such test of environmental bias.  
%%%%%%%%%%%%%%%%%%%%%%%%%%%%%%%%%%%%%%
%%%%%%%%%%%%%%%%%%%%%%%%%%%%%%%%%%%
For each cluster member we have plotted residuals from the TF relation as a function of projected distance from the nominal cluster center.  Panel (a) displays the data for all cluster members, while panels (b) and (c) differentiate between Abell cluster richness.  All data are corrected for peculiar velocity and morphological offsets.  We see no apparent trend with projected cluster radius in any of the panels.  This is in agreement with the work of Biviano \etal\ (1990) and G97b, which showed no change in the TF relation for different environments.  We consequently do not consider environmental bias to be a serious concern with our data.

%\subsection{Results}
\subsection{The Template Relation}
\label{sec:results}

%\subsubsection{Template Fitting Procedures}

Let $a_k$ be the TF zero point for cluster $k$ and $\Delta y_{{\rm pec},k} = a_k-a_{\rm tf}$ be the shift due to the cluster's peculiar motion.  The $i^{\rm th}$ cluster member's absolute magnitude can then be expressed as
\be
y_i=y_{{\rm cor},i} - \Delta y_{\rm icb,i} - \Delta y_{{\rm pec},k}
\label{eq:y}
\ee
where $\Delta y_{\rm icb}$ is the cluster's incompleteness bias correction described in Section \ref{sec:CI} and $y_{{\rm cor},i}=m_{\rm cor}-5\log_{10} ({\rm c}z_{\rm clus}/100)-25$, $m_{\rm cor}$ being the corrected apparent magnitude given by Equation 1 in Paper II and ${\rm c}z_{\rm clus}$ is the cluster systemic recessional velocity in units of \kms.  We compute the $k^{\rm th}$ cluster's zero point by averaging over the individual cluster members, i.e.
\be
a_k = {\sum_i^{N_k} (y_i-bx_i)/\epsilon_i^2 \over \sum_i^{N_k} 1/\epsilon_i^2}, \;\;
\epsilon_i^2 = (\epsilon_{x,i} b)^2 + \epsilon_{y,i}^2 + \epsilon_{\rm int}^2
\ee
with $N_k$ representing the number of cluster members in the $k^{\rm th}$ cluster.\footnote{Though we are only solving for the relation's offset here, this mode of calculation is similar in spirit to the bivariate calculations described in G97b, in that both magnitude and velocity width errors are considered.}  The uncertainty $\epsilon_{a,k}$ of the above zero point computation is described by Equation \ref{eq:epsilon_a}.  The calculation of the template zero point $a_{\rm tf}$ follows.  Alternative estimators of $a_{\rm tf}$ are:
\be
a_{\rm tf} = {\sum_k N_k a_k \over \sum_k N_k}        \;\; {\rm or} \;\;
{\sum_k a_k/\epsilon_{a,k}^2 \over \sum_k 1/\epsilon_{a,k}^2}  \;\; {\rm or} \;\;
{\sum_j^N (y_j-bx_j)/\epsilon_j^2 \over \sum_j^N 1/\epsilon_j^2}.
\ee 
The first two estimators are averages over the individual cluster zero points, weighted either by the number of cluster members or by the cluster zero point uncertainties.  The third calculation is a simple average over all $N$ cluster galaxies, with each galaxy weighted by its total uncertainty.  The three estimators give global zero points that agree within the estimated errors (0.02 magnitudes); we adopt the third computation.  Our definition of a TF template zero point is based on the assumption that the average peculiar velocity of the cluster set is null.

%\subsubsection{The Template Relation}
Table \ref{tab:results} lists results for several subsets of the data and for two different Shechter LFs: $\alpha=-0.50, M^*=-21.6$ and $\alpha=-0.75, M^*=-22.0$.  A few fit parameters are listed, including the number of galaxies used, the TF template zero point and its associated total uncertainty (i.e. considering both statistical concerns and the kinematic uncertainty described by Equation \ref{eq:sigma_a}), an estimate of the scatter, and chi squared divided by the number of degrees of freedom.
%%%%%%%%%%%%%%%%%%%%%%%%%%%%%%
\begin{table}
\caption[Template Fit Parameters]{Template Fit Parameters}
{\small \def\ea{$\epsilon_a$}
\def\sa{$\sigma_a$}
\def\dcz{$|{\rm c}z - {\rm c}z_{\rm clus}| < 1000$ \kms}
\def\c2{$\chi^2/{\rm dof}$}
% from programs/all.f

\begin{center}
\begin{tabular}{lccccc} \hline \hline
Sample                           & $N$ &  $a$   & \ea  &  \sa & \c2 \\
\hline
no incompleteness bias           & 441 & -20.96 & 0.02 & 0.39 & 1.1 \\
                                 &     &        &      &      &     \\
$\alpha_{\rm LF}=-0.50, M^*=-21.6$&    &        &      &      &     \\
                                 & 441 & -20.91 & 0.02 & 0.38 & 1.1 \\
2.5$\sigma$ clip                 & 425 & -20.91 & 0.02 & 0.34 & 0.9 \\
$60^\circ < i < 80^\circ$        & 252 & -20.91 & 0.02 & 0.38 & 1.2 \\
$i > 45^\circ$                   & 429 & -20.91 & 0.02 & 0.38 & 1.1 \\
$W > 150, y < -18.5$             & 428 & -20.92 & 0.02 & 0.37 & 1.1 \\
$T \leq Sb$                      & 139 & -20.95 & 0.03 & 0.38 & 1.3 \\
$T > Sb$                         & 302 & -20.89 & 0.02 & 0.37 & 1.0 \\
${\rm c}z > 10,000$ \kms         & 325 & -20.92 & 0.02 & 0.38 & 1.1 \\
$N_{\rm mem} \geq 5$             & 413 & -20.92 & 0.02 & 0.38 & 1.1 \\
$\theta_{\rm proj} < $1 Abell radius      & 347 & -20.91 & 0.02 & 0.39 & 1.1 \\
\dcz                             & 368 & -20.91 & 0.02 & 0.38 & 1.1 \\
includ. foreground \& background & 510 & -20.90 & 0.02 & 0.39 & 1.2 \\
                                 &     &        &      &      &     \\
$\alpha_{\rm LF}=-0.75, M^*=-22.0$&    &        &      &      &     \\
                                 & 441 & -20.90 & 0.02 & 0.38 & 1.1 \\
2.5$\sigma$ clip                 & 425 & -20.90 & 0.02 & 0.34 & 0.9 \\
$60^\circ < i < 80^\circ$        & 252 & -20.90 & 0.02 & 0.38 & 1.2 \\
$i > 45^\circ$                   & 429 & -20.90 & 0.02 & 0.37 & 1.1 \\
$W > 150, y < -18.5$             & 427 & -20.91 & 0.02 & 0.37 & 1.1 \\
$T \leq Sb$                      & 139 & -20.95 & 0.03 & 0.38 & 1.3 \\
$T > Sb$                         & 302 & -20.88 & 0.02 & 0.37 & 0.9 \\
${\rm c}z > 10,000$ \kms         & 325 & -20.91 & 0.02 & 0.38 & 1.1 \\
$N_{\rm mem} \geq 5$             & 413 & -20.91 & 0.02 & 0.34 & 1.1 \\
$\theta_{\rm proj} < $1 Abell radius      & 347 & -20.90 & 0.02 & 0.38 & 1.1 \\
\dcz                             & 368 & -20.90 & 0.02 & 0.38 & 1.1 \\
includ. foreground \& background & 510 & -20.89 & 0.02 & 0.39 & 1.2 \\
\hline
\end{tabular}
\end{center}
 }
\label{tab:results}
\end{table}
%%%%%%%%%%%%%%%%%%%%%%%%%%%%%
Notice that the dispersion and the zero point uncertainty remain relatively unchanged across all subsets.  The fact that the zero point differs slightly for late and early type galaxies is expected -- we saw earlier that the morphological type correction of $-0.32$ magnitudes advocated in G97b for Sa and Sab galaxies is possibly too large for our sample by a few hundredths of a magnitude.  We should also point out that the slightly brighter zero point seen for objects beyond 10,000 \kms\ works against recent claims by Tammann (1998) and Zehavi (1998) that the Hubble constant is higher within \about 10,000 \kms\ than it is beyond this distance.  For our purposes, a fractional decrease by $\epsilon_H$ in $H_{\circ}$ beyond 10,000 \kms\ should yield a zero point that is {\it fainter} by $\epsilon_H 5\log_{10}$$e$, or 0.04 mags for $\epsilon_H=0.02$.  In contrast, our data reflect a slightly {\it brighter} zero point beyond 10,000 \kms.

\begin{figure}[!ht]
%\clearpage
\centerline{\psfig{figure=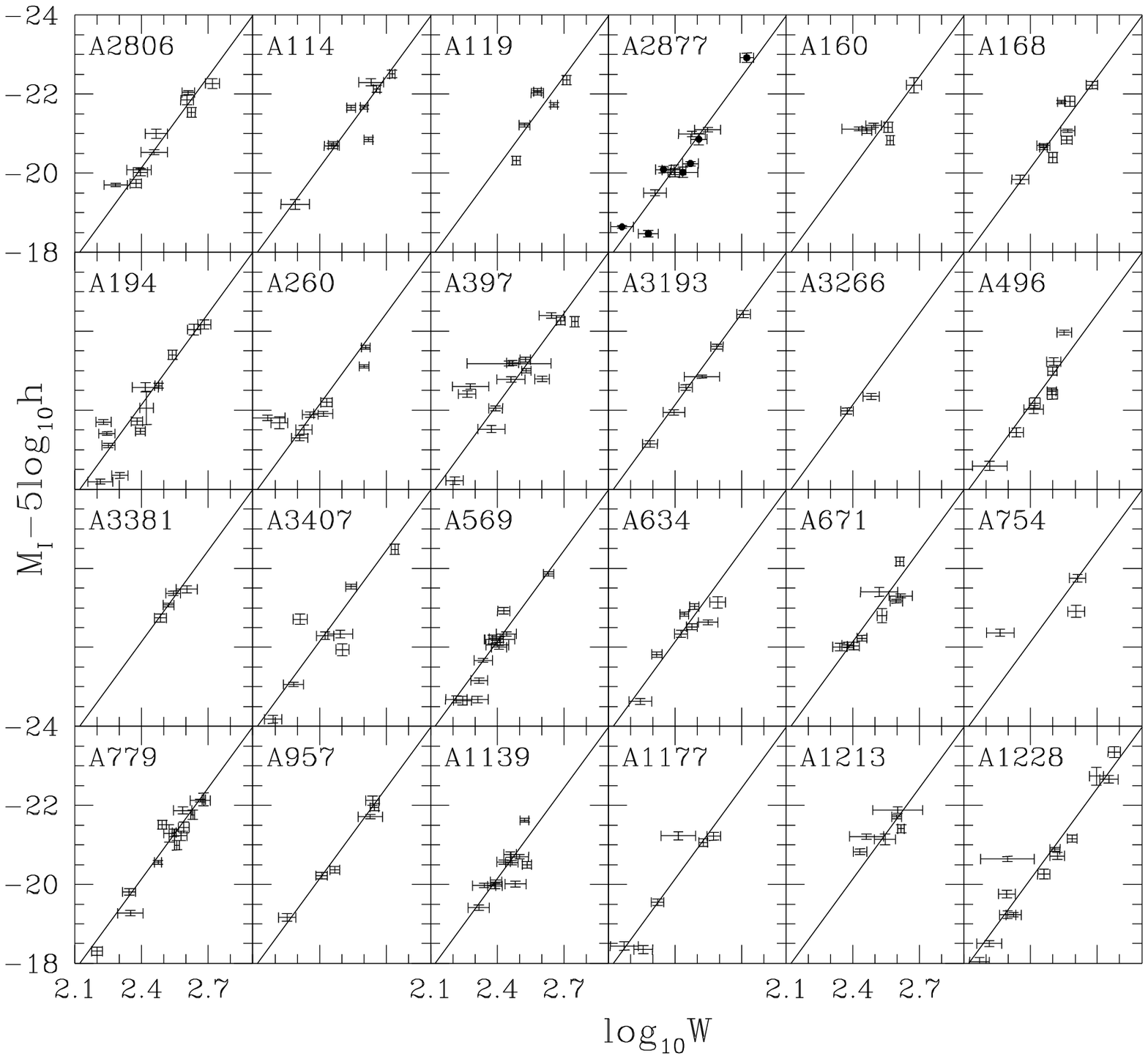,width=5.5in,bbllx=52pt,bblly=220pt,bburx=570pt,bbury=700pt}}
\caption[Cluster Tully-Fisher Diagrams]
{\ TF plots for the clusters are given.  Only cluster members are included in these plots.  Each luminosity is corrected for cluster incompleteness and cluster peculiar velocity.  The morphological offset described in Section \ref{sec:morph} for Sb and earlier type galaxies is incorporated.  The new TF template relation is drawn as a reference (Equation \ref{eq:TF}).}
\label{fig:TFcor}
\end{figure}

Figure \ref{fig:TFcor} gives the TF plots for each cluster and Figure \ref{fig:TFall_cor} combines the data for all SCII galaxies.  The data are corrected for cluster incompleteness bias and cluster peculiar motion, in accordance with Equation \ref{eq:y}.  In the A2877, A1736, A1983, and A2295 panels, the error bars containing filled circles represent members of ``A2877b,'' ``A1736b,'' ``A1983b,'' and ``A2295b,'' respectively.  The new TF template is drawn as well:
\be
M_I-5\log h = -7.68 (\log W -2.5) - 20.91 \;\; {\rm mag}.
\label{eq:TF}
\ee
%%%%%%%%%%%%%%%%%%%%%%%%%%%%%%%%%%%%%%
\begin{figure}[!ht]
\centerline{Figure \ref{fig:TFcor} (Continued)}
\centerline{\psfig{figure=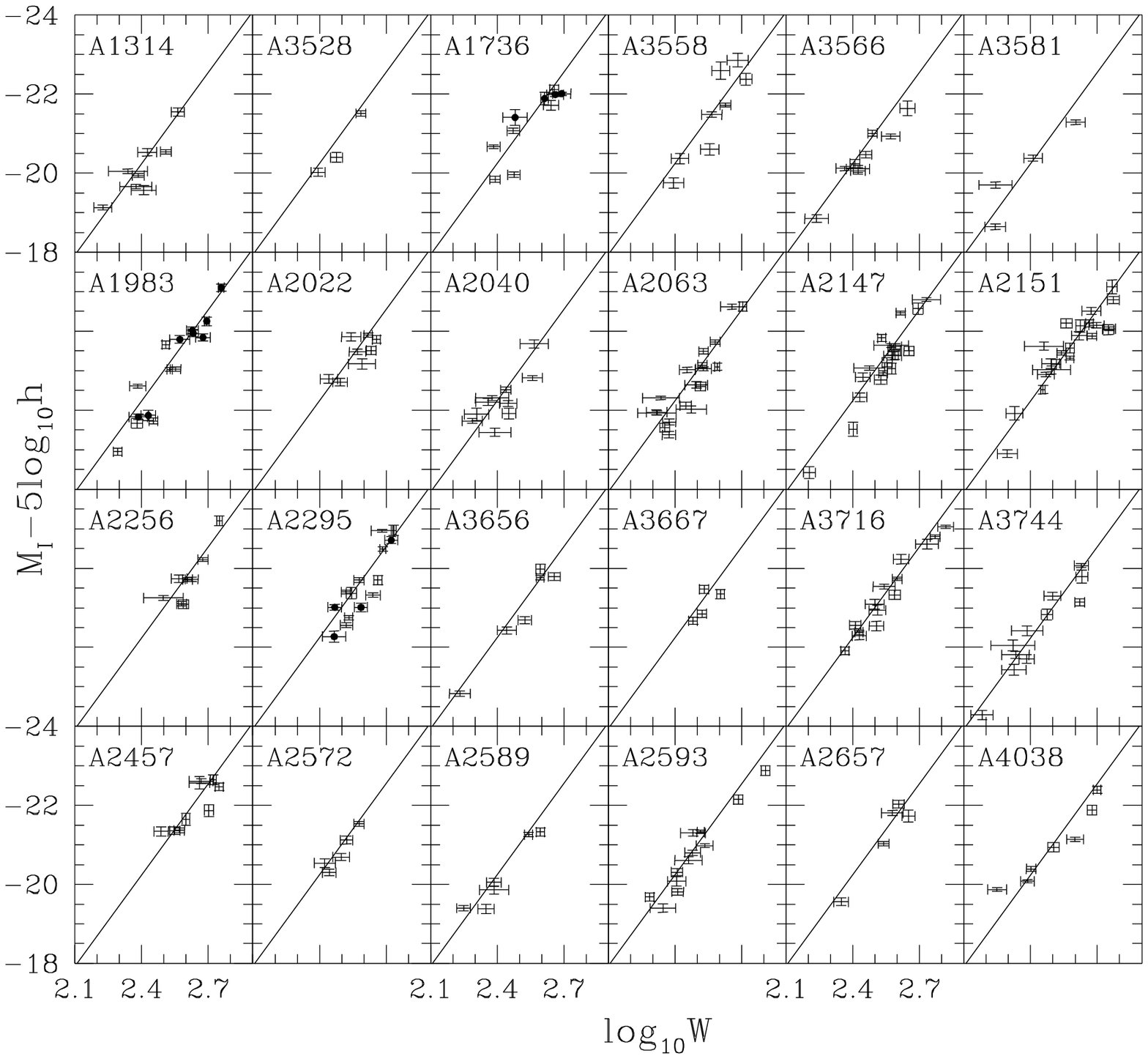,width=5.5in,bbllx=52pt,bblly=220pt,bburx=570pt,bbury=700pt}}
\end{figure}
%%%%%%%%%%%%%%%%%%%%%%%%%%%%%%%%%%%
 
%%%%%%%%%%%%%%%%%%%%%%%%%%%%%%%%%%%%%%
\begin{figure}[ht]
\centerline{\psfig{figure=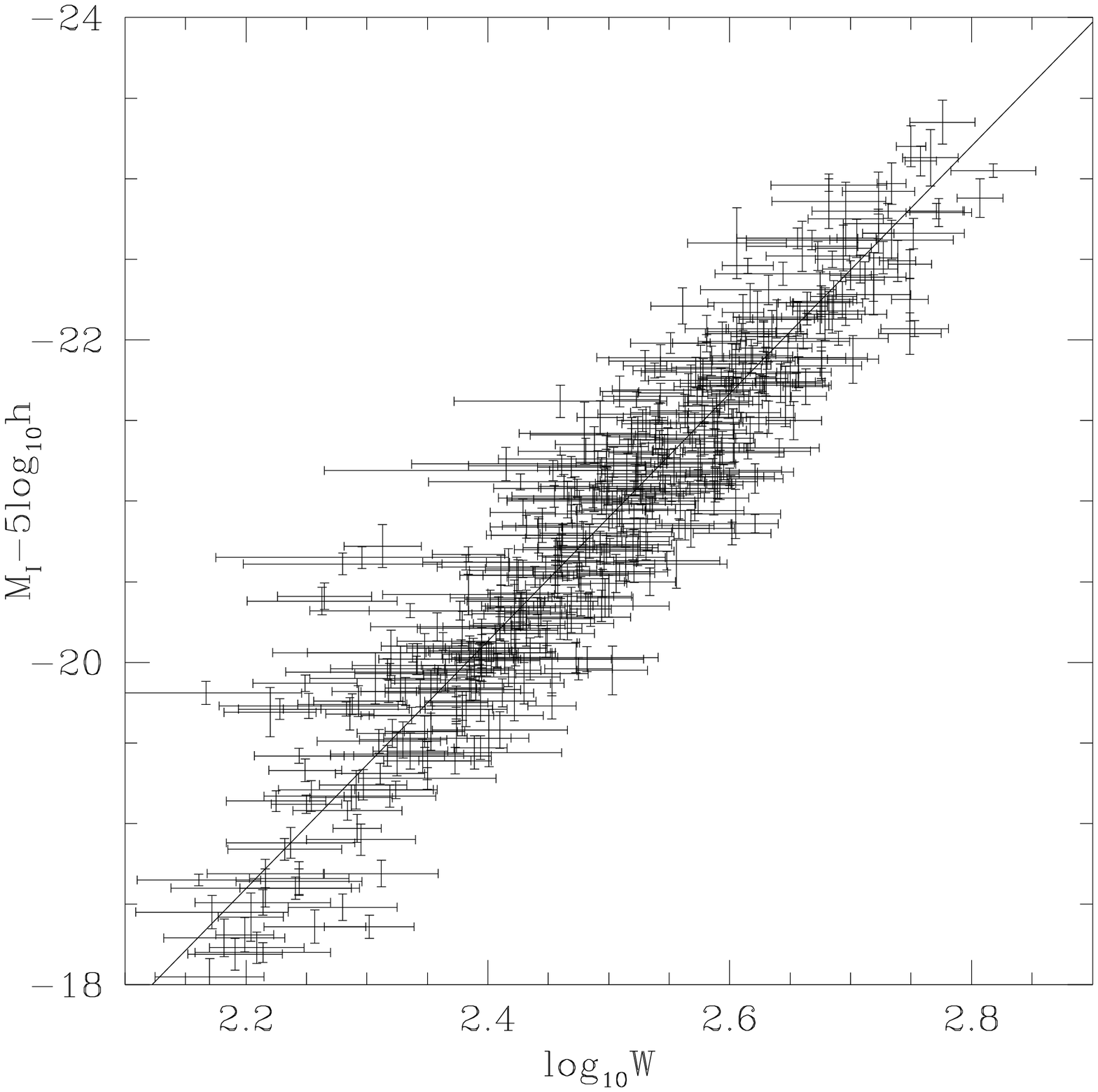,width=4.0in,bbllx=13pt,bblly=150pt,bburx=566pt,bbury=696pt}}
\caption[Bias-Corrected Tully-Fisher Data]
{\ Plotted are the TF data for all galaxies, corrected for peculiar motions, incompleteness bias, and the morphological type offset described in Section \ref{sec:morph}.  The new TF template relation is drawn as a reference (Equation \ref{eq:TF}).}
\label{fig:TFall_cor}
\end{figure}
%%%%%%%%%%%%%%%%%%%%%%%%%%%%%%%%%%%

\begin{figure}[ht]
\centerline{\psfig{figure=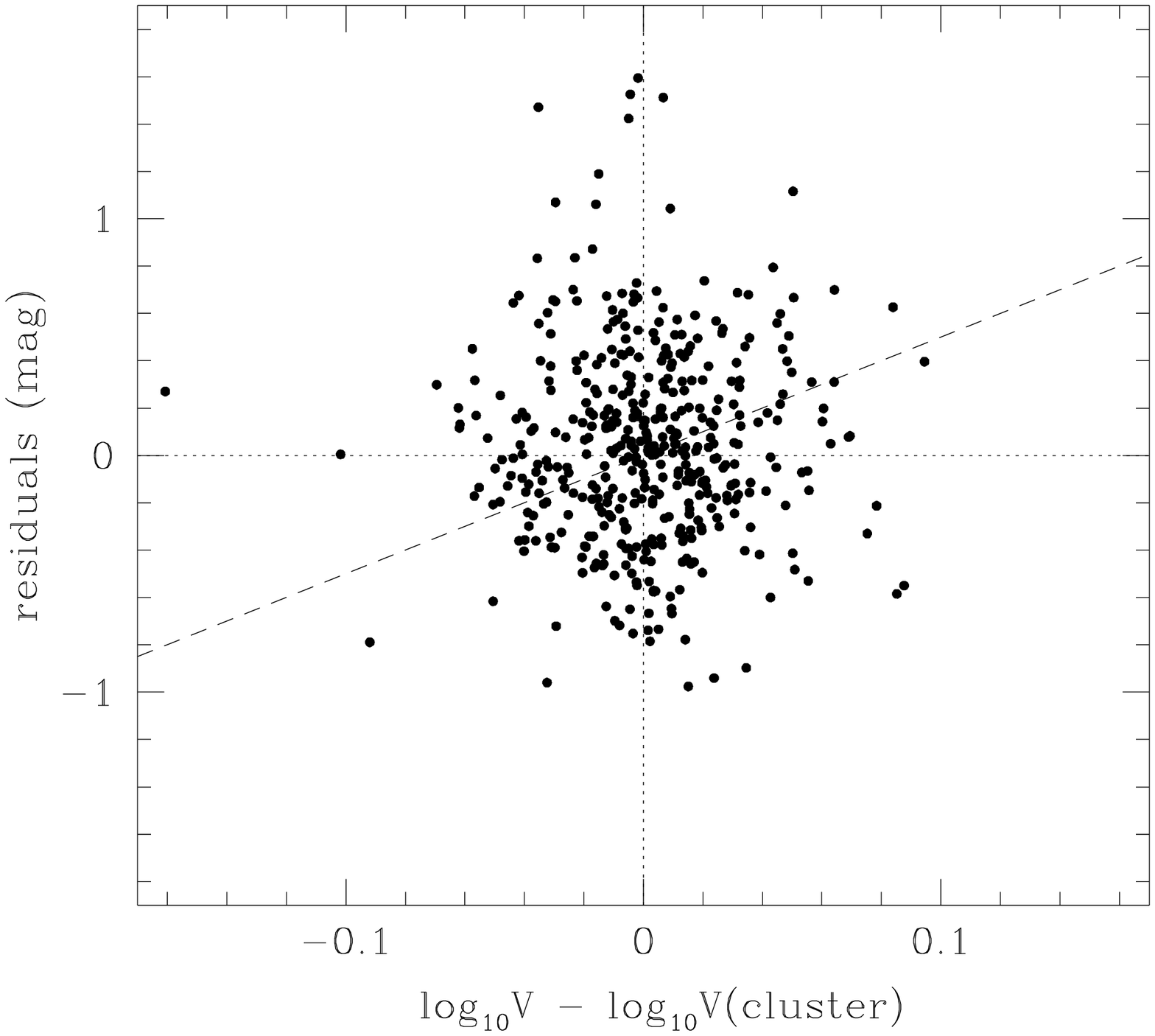,width=3.0in,bbllx=20pt,bblly=205pt,bburx=566pt,bbury=690pt}}
\caption[A Test of Cluster Membership I]
{\ A test of cluster membership is displayed in the form of TF residuals as a function of separation from the cluster redshift.  A line of slope 5 is given, the relation on which field galaxies incorrectly assumed to be cluster members should lie.}
\label{fig:czresid_all}
\end{figure}

The residuals given in Figure \ref{fig:czresid_all} indicate the quality of the cluster membership assignments.  
%%%%%%%%%%%%%%%%%%%%%%%%%%%%%%%%%%%%%%
%%%%%%%%%%%%%%%%%%%%%%%%%%%%%%%%%%%
Since the abscissa is the difference between individual and cluster redshifts, the center of each cluster corresponds to the point (0,0).  Moreover, residuals from field galaxies that are receding according to Hubble expansion but were incorrectly declared cluster members should, on average, lie on the line of slope 5.  There is no clear evidence for a significantly large subsample of improperly assigned memberships.

\section{The Peculiar Velocity Sample}
\begin{figure}[!ht]
\centerline{\psfig{figure=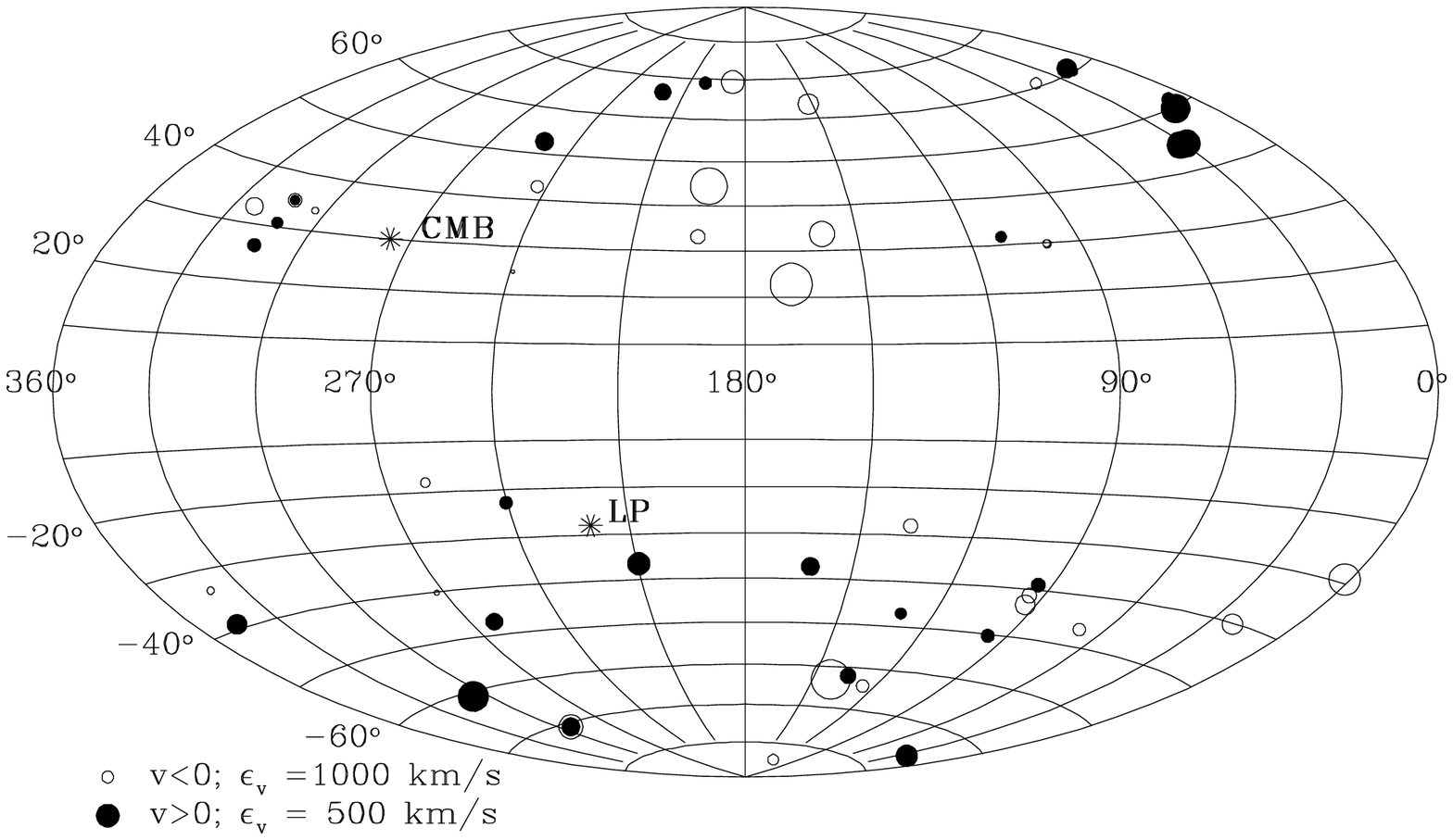,width=5.5in,bbllx=30pt,bblly=234pt,bburx=585pt,bbury=560pt}} % use vpec/allsky2.f
\caption[Aitoff Projection of the Peculiar Velocities]
{\ The all-sky distribution of the SCII peculiar velocity sample in Galactic coordinates.  The symbol diameters are inversely proportional to the peculiar velocity uncertainties.  The two examples in the lower left give the scale.  Filled (open) circles refer to positive (negative) peculiar velocities in the CMB frame.  Asterisks mark the apices of the motion of the Local Group with respect to the CMB and the LP cluster inertial frame.}
\label{fig:aitoff_Vpec}
\end{figure}

We interpret the departure of a cluster's average zero point from that of the template relation as an indication of peculiar motion, with larger departures from the template implying larger amplitude peculiar velocities.  Quantitatively, for a cluster at a redshift $z$ with an average departure from the template of $a-a_{\rm tf}$, we write the peculiar velocity as 
\be
V = {\rm c}z \big( 1 - 10^{0.2(a-a_{\rm tf})}).
\label{eq:V}
\ee
We compute peculiar velocities and a measure of their errors for the SCII cluster sample and display the results in Table \ref{tab:vpec} and Figure \ref{fig:aitoff_Vpec}, an Aitoff projection of Galactic coordinates.
%%%%%%%%%%%%%%%%%%%%%%%%%%%%%%
\begin{table}[!ht]
% notice I switched the definitions of the bounding box... 
% 4.2in is biggest width allowable
%\centerline{\psfig{figure=tables/7.1.ps,width=4.2in,bbllx=422pt,bblly=689pt,bburx=55pt,bbury=13pt}}
\caption[Cluster Peculiar Velocities]{Cluster Peculiar Velocities}
{\small \def\hel{${\rm c}z_{\rm hel}$}
\def\lg{${\rm c}z_{\rm lg}$}
\def\cmb{${\rm c}z_{\rm cmb}$}
\def\vtf{$V_{\rm tf,cmb}$}
\def\vp{$V_{\rm pec,cmb}$}

\begin{center}
\begin{tabular}{lrrrrrr} \hline
Cluster & $a-a_{\rm tf}$ &  \hel &   \lg &  \cmb &  \vtf &     \vp     \\
\hline
A2806  & -0.135(112) &  8019 &  7918 &  7867 &  7403 &   464(0382) \\
A114   &  0.071(134) & 17436 & 17487 & 17144 & 17722 &  -578(1111) \\
A119   &  0.053(153) & 13470 & 13607 & 13141 & 13416 &  -275(0988) \\
A2877  &  0.039(155) &  7165 &  7097 &  6974 &  7078 &  -104(0489) \\
A2877b & -0.030(158) &  9231 &  9163 &  9040 &  8733 &   307(0634) \\
A160   & -0.042(176) & 12390 & 12577 & 12072 & 11792 &   280(0977) \\
A168   & -0.106(127) & 13365 & 13496 & 13049 & 12370 &   679(0725) \\
A194   &  0.089(127) &  5343 &  5460 &  5037 &  5253 &  -216(0302) \\
A260   &  0.218(162) & 10925 & 11135 & 10664 & 11839 & -1175(0835) \\
A397   & -0.144(149) &  9803 &  9916 &  9594 &  9041 &   553(0630) \\
A3193  & -0.100(144) & 10558 & 10381 & 10522 & 10072 &   450(0668) \\
A3266  &  0.392(248) & 17774 & 17564 & 17782 & 20482 & -2700(2345) \\
A496   & -0.128(123) &  9860 &  9779 &  9809 &  9243 &   566(0513) \\
A3381  & -0.157(176) & 11410 & 11196 & 11510 & 10712 &   798(0868) \\
A3407  &  0.090(221) & 12714 & 12450 & 12861 & 13040 &  -179(1235) \\
A569   &  0.061(099) &  5927 &  5998 &  6011 &  6168 &  -157(0280) \\
A634   &  0.088(125) &  7822 &  7917 &  7922 &  8144 &  -222(0469) \\
A671   &  0.124(118) & 15091 & 15046 & 15307 & 15427 &  -120(0838) \\
A754   &  0.047(414) & 16282 & 16052 & 16599 & 16691 &   -92(3294) \\
A779   &  0.026(095) &  6967 &  6929 &  7211 &  7311 &  -100(0320) \\
A957   &  0.119(144) & 13464 & 13261 & 13819 & 14685 &  -866(0974) \\
A1139  & -0.107(118) & 11851 & 11665 & 12216 & 11522 &   694(0629) \\
A1177  & -0.008(150) &  9755 &  9661 & 10079 & 10028 &    51(0689) \\
A1213  & -0.180(144) & 14006 & 13952 & 14304 & 13560 &   744(0899) \\
A1228  &  0.109(099) & 10517 & 10489 & 10794 & 11397 &  -603(0517) \\
A1314  &  0.034(125) &  9764 &  9815 &  9970 & 10104 &  -134(0582) \\
A3528  &  0.201(203) & 16458 & 16224 & 16770 & 18211 & -1441(1703) \\
A1736  &  0.010(176) & 10397 & 10186 & 10690 & 10739 &   -49(0887) \\
A1736b & -0.039(176) & 13724 & 13513 & 14017 & 13831 &   186(1121) \\
A3558  & -0.107(152) & 14343 & 14122 & 14626 & 13948 &   678(0981) \\
A3566  &  0.074(118) & 15370 & 15147 & 15636 & 15400 &   236(0837) \\
A3581  &  0.042(197) &  6865 &  6681 &  7122 &  7261 &  -139(0659) \\
A1983b & -0.277(125) & 11332 & 11340 & 11524 & 10233 &  1291(0589) \\
A1983  & -0.057(192) & 13526 & 13537 & 13715 & 13286 &   429(1165) \\
A2022  &  0.127(125) & 17262 & 17327 & 17412 & 18546 & -1134(1067) \\
A2040  &  0.001(135) & 13440 & 13429 & 13616 & 13404 &   212(0839) \\
A2063  & -0.158(085) & 10444 & 10448 & 10605 &  9925 &   680(0398) \\
A2147  & -0.060(091) & 10493 & 10558 & 10588 & 10285 &   303(0427) \\
A2151  & -0.074(085) & 11005 & 11079 & 11093 & 10781 &   312(0424) \\
A2256  &  0.008(125) & 17442 & 17673 & 17401 & 17345 &    56(0998) \\
A2295b &  0.042(177) & 18701 & 18945 & 18633 & 19041 &  -408(1587) \\
A2295  &  0.163(123) & 24622 & 24868 & 24554 & 25699 & -1145(1448) \\
A3656  &  0.011(144) &  5750 &  5746 &  5586 &  5658 &   -72(0375) \\
A3667  &  0.417(176) & 16581 & 16491 & 16477 & 19511 & -3034(1582) \\
A3716  & -0.053(095) & 13764 & 13701 & 13618 & 13259 &   359(0581) \\
A3744  &  0.021(109) & 11386 & 11465 & 11123 & 11273 &  -150(0578) \\
A2457  &  0.004(118) & 17643 & 17845 & 17280 & 17424 &  -144(0946) \\
A2572  & -0.079(158) & 11857 & 12106 & 11495 & 11059 &   436(0803) \\
A2589  &  0.048(144) & 12289 & 12531 & 11925 & 12119 &  -194(0804) \\
A2593  &  0.140(103) & 12415 & 12651 & 12049 & 12810 &  -761(0605) \\
A2657  & -0.007(158) & 12028 & 12239 & 11662 & 11630 &    32(0844) \\
A4038  & -0.022(134) &  9012 &  9062 &  8713 &  8645 &    68(0534) \\
\hline					      
\end{tabular}
\end{center}
 }
\label{tab:vpec}
\end{table}
%%%%%%%%%%%%%%%%%%%%%%%%%%%%%
%%%%%%%%%%%%%%%%%%%%%%%%%%%%%%%%%%%%%%
%%%%%%%%%%%%%%%%%%%%%%%%%%%%%%%%%%%
The symbols plotted in the figure reflect both the radial directions of the peculiar velocities and the strengths of the measurements -- in the CMB reference frame, open (filled) circles represent approaching (receding) clusters and the circle size is inversely proportional to the accuracy of the measurement.  The largest cluster peculiar velocities, e.g. those for A3266 and A3667, are also the most uncertain, as the clusters are poorly sampled.  

Our sample includes the central portions of various high density peaks and/or superclusters in the local Universe.  It is significant to note that their motions are consistent with small departures from rest in the CMB frame, within the quoted error.  For instance, the Shapley Supercluster, represented here by its core (A3558) and some peripheral members (A1736, A3528, and A3566), has an average CMB velocity of $118\pm495$ \kms.  The Hercules region (A2147 and A2151) and the A2572/2589/2593/2657 supercluster are also slow movers, with average peculiar velocities of $307\pm301$ \kms\ and $-222\pm372$ \kms, respectively.  These small motions are consistent with the notion that such massive systems best represent ``kinematic anchors'' in the local velocity field.

\subsection{The One-Dimensional Peculiar Velocity Distribution}

\begin{figure}[!ht]
\centerline{\psfig{figure=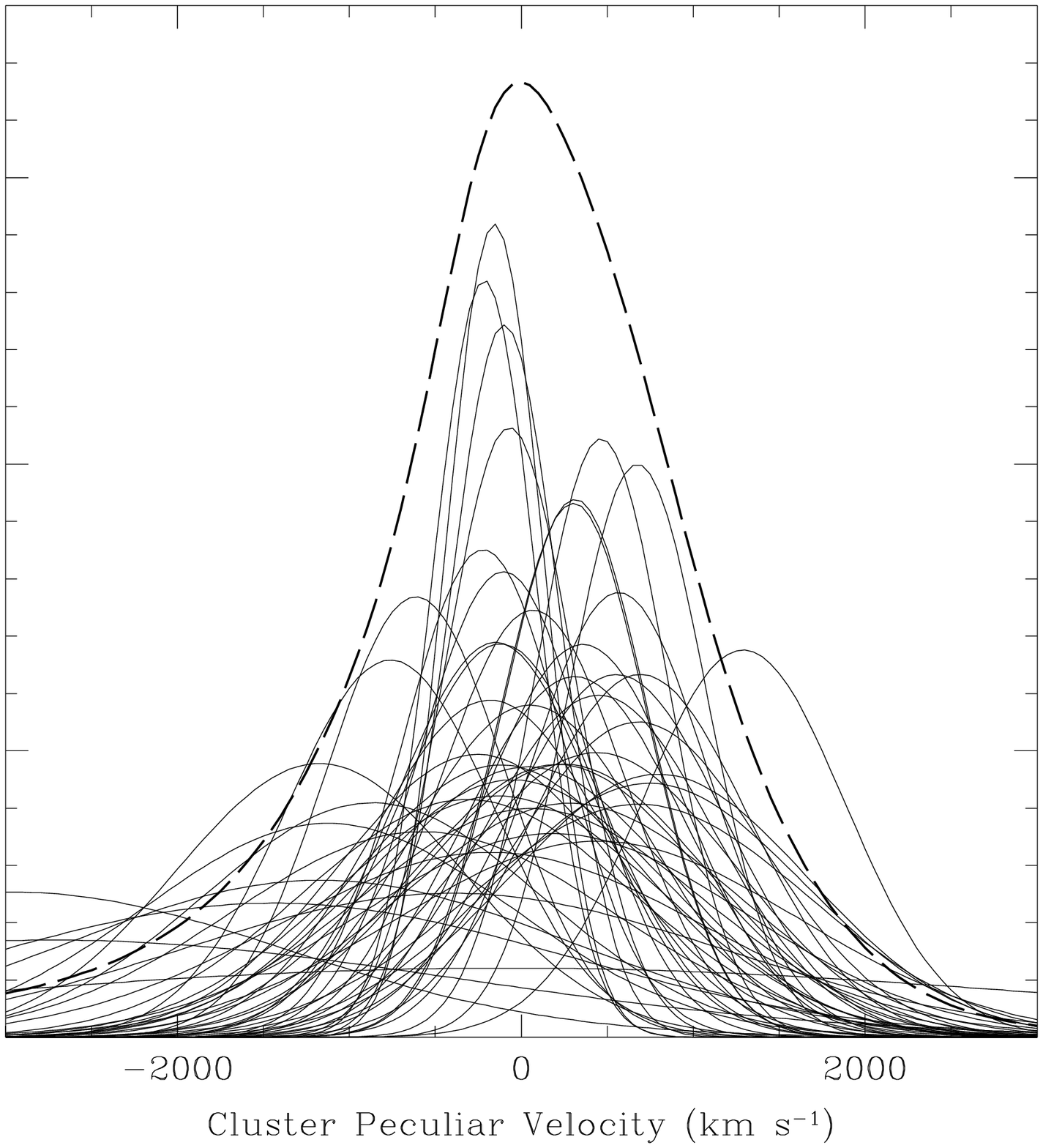,width=4.0in,bbllx=70pt,bblly=155pt,bburx=572pt,bbury=695pt}}
\caption[Peculiar Velocity Distribution]
{\ A display of equal area Gaussians representing the peculiar velocity sample.  Each Gaussian is centered at the value of a cluster peculiar velocity, has a dispersion given by the uncertainty in the peculiar velocity measurement, and has an amplitude that is inversely proportional to the uncertainty.  The sum of the Gaussian profiles is given by the thick dashed line.}
\label{fig:gauss_vpecs}
\end{figure}

It is useful to estimate the line-of-sight distribution of peculiar velocities.  The amplitude of that distribution has been known to be a very sensitive discriminator of cosmological models (see, for example, Bahcall \& Oh 1996).  The SCII cluster sample is relatively distant and the cluster membership counts are relatively anemic when compared to the SCI sample of G97a.  Consequently, SCII peculiar velocities are much less certain and the overall distribution shown in Figure \ref{fig:gauss_vpecs} is significantly broadened by measurement errors.  The peculiar velocities are represented by equal area Gaussians centered at the peculiar velocity of each cluster with dispersions equal to the estimated peculiar velocity errors.  The thick dashed line superimposed on the plot is the sum of the individual Gaussians (its amplitude has been rescaled for plotting purposes).
%%%%%%%%%%%%%%%%%%%%%%%%%%%%%%%%%%%%%%
%%%%%%%%%%%%%%%%%%%%%%%%%%%%%%%%%%%
The 1$\sigma$ dispersion in the observed distribution of peculiar velocities is found from a Gaussian fit to the dashed line: $\sigma_{1{\rm d,obs}}=796$ \kms.  This value, however, is biased high by measurement errors.  Recovering an estimate of the true value can easily be obtained via Monte Carlo simulations, yielding $\sigma_{\rm 1d} = 341\pm93$ \kms where the error estimate derives from the scatter in the dispersions of the simulated samples.  This value of $\sigma_{\rm 1d}$ is consistent with a relatively low density Universe (Giovanelli \etal\ 1998b; Bahcall \& Oh 1996; Borgani \etal\ 1997; Watkins 1998; Bahcall, Gramann \& Cen 1994a; Croft \& Efstathiou 1994).

\section{Summary}
We have presented TF data and estimated the peculiar velocities of 52 rich Abell clusters spread across  the sky and distributed between \about 50 and 200\h\ Mpc.  Optical rotation curves and $I$ band photometry for 522 spiral galaxies in the fields of these systems have been obtained and presented in separate publications.

In conjunction with the robust TF slope extracted from the relatively nearby SCI cluster sample of Giovanelli and coworkers, we find the $I$ band TF relation to follow
\be
M_I - 5\log_{10}h = -7.68 (\log_{10}W-2.5) - 20.91 \; {\rm mag}.
\ee
The relation has an average scatter of 0.38 magnitudes.  The zero point of the TF template has a statistical accuracy of 0.02 mag; combined with a kinematical uncertainty of 0.01 mag, which is limited by the assumption that the 52 clusters' average peculiar velocity is null, the overall uncertainty of the TF zero point is 0.02 mag.  

Peculiar velocities are obtained for each of the 52 clusters, with reference to the global template relation.  The typical uncertainty on the peculiar velocity of each cluster is $\pm$0.06c$z$, where c$z$ is the mean cluster velocity.  The rms line-of-sight component of the cluster peculiar velocity for our sample, debroadened for measurement errors, is $341\pm93$ \kms.  This number agrees with that determined for the SCI sample.

\acknowledgements
The results presented here are based on observations carried out at the Palomar Observatory (PO), at the Kitt Peak National Observatory (KPNO), at the Cerro Tololo Inter--American  Observatory (CTIO), and the Arecibo Observatory, which is part of the National Astronomy and Ionosphere Center (NAIC).  KPNO and CTIO are operated by Associated Universities for Research in Astronomy and NAIC is operated by Cornell University, all under cooperative agreements with the National Science Foundation.  The Hale telescope at the PO is operated by the California Institute of Technology under a cooperative agreement with Cornell University and the Jet Propulsion Laboratory.  This research was supported by NSF grants AST94-20505 and AST96--17069 to RG and AST95-28960 to MH.  LEC was partially supported by FONDECYT grant \#1970735.
 
%\newpage

%\input{appendix.tex}
\end{document}